\documentclass[aps,showpacs,pre,twocolumn,groupedaddress]{revtex4-1}

\usepackage{amssymb}
\usepackage{amsmath}
\usepackage{amsthm}
\usepackage{graphicx}
\usepackage{dcolumn}
\usepackage{bm}

\begin{document}


\title{Hierarchy and Polysynchrony in an adaptive network}


\author{V. Botella-Soler}
\email{vbsoler@ist.ac.at} \affiliation{IST Austria (Institute of Science and Technology Austria), \\ Am Campus 1, A-3400 Klosterneuburg, Austria}
\author{P. Glendinning}
\email{p.a.glendinning@manchester.ac.uk}
 \affiliation{School of Mathematics and\\
 Centre for Interdisciplinary Computational and Dynamical Analysis (CICADA),\\
 University of Manchester, Manchester M13 9PL, U.K.}



\begin{abstract}
We describe a simple adaptive network of coupled chaotic maps.
The network reaches a stationary state (frozen topology) for
all values of the coupling parameter, although the dynamics of the maps at the nodes
of the network can be non-trivial. The structure of the
network shows interesting hierarchical properties and in
certain parameter regions the dynamics is polysynchronous:
nodes can be divided in differently synchronized classes but
contrary to cluster synchronization, nodes in the same class
need not be connected to each other. These complicated
synchrony patterns have been conjectured to play roles in
systems biology and circuits. The adaptive system we study
describes ways whereby this behaviour can evolve from
undifferentiated nodes.
\end{abstract}


\maketitle


\section{Introduction}

Networks allow us to model a huge variety of complex systems where a multitude of agents dynamically interact  \cite{boccaletti2006complex,
newman2006structure}. The agents are modeled as nodes and the links of the network stand for their interactions. When the dynamics of the agents can affect the pattern of interactions, i.e. change the structure of the network, we speak of complex adaptive networks \cite{Gross2008, gross2009adaptive}. These networks can show a variety of dynamical and structural properties depending on the dynamics of the agents, the nature of the interactions or the adaptation mechanism \cite{Ito2001, Ito2003, fan2004evolving, Berg2004, Gong2004,
Zhou2006, lu2007adaptive, aoki2009co, Li2010}. Adaptive networks have been already applied to different problems such as neural networks \cite{kwok2007robust, meisel2009adaptive, gomez2009adaptive,
gleiser2010modelling}, epidemic spreading \cite{Gross2006,gross2008robust,shaw2010enhanced} and opinion formation \cite{kozma2008consensus, nardini2008s}.

The dynamics of the agents at the nodes of adaptive networks can be very complicated. In \cite{bg2011polyletter} we described
numerical simulations of an adaptive network that could evolve into a state with polysynchronous dynamics at appropriate parameter values.
Polysynchrony is a form of network synchronization where groups of nodes synchronize without being directly connected \cite{stewart2003symmetry,golubitsky2004some,field2004combinatorial,
aguiar2009dynamics, agarwal2010dynamical}. The term sublattice synchronization has also been used for this same phenomenon \cite{Kestler2007, Kestler2008, Kanter2011}.
The aim of this paper is to provide a more rigorous and complete analysis of the adaptive network introduced in \cite{bg2011polyletter}.
We describe in detail the dynamical regimes this model of adaptive network can show and explain the different regimes through the analytical study of the stability of the different attractors or synchronized states. We also prove results about the asymptotic stationarity of the network topology
and describe the hierarchical nature of this frozen state (although here we add one simplifying rule). The structure of the paper is as follows. In section \ref{sec:model} the adaptive
network introduced in \cite{bg2011polyletter} is defined. The dynamics of the network topology has a stochastic element driven by a
homophilic principle, so nodes in similar states `like' to be connected together. At each time step the network topology can change according to
 a set of rules, thus changing the inputs to the dynamics at the nodes. We refer to this process as \emph{rewiring}. In section \ref{sec:results}
 we explore the dynamics of the network numerically. We show that the network reaches a frozen state where the rewiring stops. The transient times to the frozen state are evaluated as a function of the network size and the coupling strength.  The different dynamical regimes are also described in this section in terms of the synchronization of the nodes. We provide several examples of polysynchronous networks and study the probability of finding polysynchrony as a function of the coupling strength. The numerical observations show that the synchronization effect is very strong, and the dynamics at different nodes
 can become indistinguishable at machine accuracy. This effect, which we believe is interesting in that it reflects what any finite measurement
 could discern, means that some of the final topological states observed are extremely unlikely from a mathematical point of view. In
 section \ref{sect:freeze} we show formally that a closely related network rule that eliminates these mathematically unlikely states
 must lead to a stationary topology. In section \ref{sec:discussion} we summarize and discuss the main results of this work and their potential applications.
 The detailed stability analysis of the fully synchronous and the polysynchronous states is given in two appendices.

 Many accounts of adaptive networks concentrate on the increased complexity of the evolving network topology (to scale-free networks for example). In
 contrast, the systems described in this paper evolve towards a stationary network topology with some striking
 features such as a strong hierarchical structure and
 polysynchronous dynamics at the nodes. Our models therefore point the way to rather different application areas: the evolution to
 networks with relatively simple structure having dynamics correlated in different nodes that are not directly connected by the network lends itself
 to interpretations in terms of functional differentiation of initially equivalent units, where the differentiated systems are distributed
 across the network rather than clustered. This and other possible applications in biological and social systems is commented on further in
 the final section of this paper.

\section{The model}\label{sec:model}
The model consists of a directed network of $N$ nodes where the
dynamics of the $i$th node ($i=1,\dots,N$) are given by
\begin{equation}
x^{i}_{n+1}=f(x^{i}_{n})+\frac{\varepsilon}{m}\sum_{j=1}^{N}A^{ij}_{n}(f(x^{j}_{n})-f(x^{i}_{n})).
\label{eq:localdyn}
\end{equation}
We choose $f$ to be the fully-chaotic logistic map
$f(x)=4x(1-x)$ and $A_n$ is the adjacency matrix of the network
at time step $n$, so $A^{ij}_{n}=k$ if there are $k$
directed edges from $j$ to $i$. In the figures we represent the directed edges by an edge with an arrow indicating the direction of the flow
of information. Thus the head of the arrow is the node that receives the input and the tail of the arrow is the node that influences the
node at the head, i.e. if $A_n^{ij}\ne 0$ there will be a directed edge (an arrow) from node $j$ to node $i$.
Each node is assigned the same fixed number $m$ of incoming links so
\begin{equation}
\sum_{j=1}^{N}A^{ij}_{n}=m, \label{eq:mneigh}
\end{equation}
and we choose $m=N-1$ throughout this paper. The input degree of the nodes is therefore fixed. This is particularly important for the interpretation of the examples we show throughout the paper where we have avoided labelling the weights of the connections; they always sum to $m$. Moreover, we will not allow a link from a node to itself so $A^{ii}_{n}=0$ for all $n$.

At each iteration the $i$th node is influenced by the dynamics
of those nodes to which it is connected by an incoming arrow.
We will call these nodes the \textit{neighbours} of node $i$.
Due to the condition imposed by (\ref{eq:mneigh}), a node can
have at most $m$ neighbours.

As indicated in the introduction, the network topology changes according to a homophilic principle. At each time step the node dynamics
evolves according to (\ref{eq:localdyn}). The values of the map $f$ at each node is compared to the values of $f$ at its neighbours and then a 'bad' set
of neighbours is identified. These are those with $f$ values far from that at the node they influence. The connections to the bad node
are then changed at random to nodes that are not bad, then the process repeats. More precisely, the nodes rewire their links through the
following mechanism. At each iteration $n$ we compute the distance matrix $D^{ij}_n$
\begin{equation}
D^{ij}_n=\left\{
\begin{tabular}{ll}
      $|f(x^{i}_{n})-f(x^{j}_{n})|$, & \ {\rm if}~$A^{ij}_{n-1}\neq0$ \\
      $0$, & \ {\rm if}~ $A^{ij}_{n-1}=0$
     \end{tabular}
     \right.\label{eq:distance}
\end{equation}
and calculate from it the mean distance of a node to all its
neighbours
\begin{equation}
\langle
D\rangle^{i}_{n}=\frac{1}{a^{i}_{n}}\sum_{j=1}^{N}D^{ij}_{n}
\end{equation}
where $a^{i}_{n}$ is the unweighted number of neighbours of node $i$ at
time step $n$, i.e. the sum over $j$ of ${\rm sign}(A^{ij}_{n-1})$.

We have choosen the rewiring to be homophily-driven, so nodes prefer
to be connected to nodes being in a similar state. Therefore, we identify
the \textit{bad} neighbours $\mathcal{B}^{i}_n$ of each
node $i$ at iteration $n$
\begin{equation}\label{eq:criterionbad}
j\in\mathcal{B}^{i}_n \quad\text{if}\quad D^{ij}_{n}>\langle
D\rangle^{i}_{n}.
\end{equation}
Thus a neighbour $j$ is considered \textit{bad} if its distance
$D^{ij}_{n}$ to the node is larger than the average distance of
the neighbourhood $\langle D\rangle^{i}_{n}$. The \textit{good}
neighbours of node $i$ are then given by
\begin{equation}\label{eq:criteriongood}
\mathcal{G}^{i}_n=\{1,\dots ,N\}\backslash\left(\mathcal{B}^{i}_n \cup\{i\}\right).
\end{equation}

Once the good and bad neighbours have been identified node $i$
will break the links coming from $\mathcal{B}^{i}_n$ and
randomly rewire them to nodes in $\mathcal{G}^{i}_n$. Let
$b^{i}_n$ be the number of \textit{bad} connections, i.e. the sum of the connections to $i$
from bad neighbours:
\begin{equation}\label{eq:bvalency}
b^{i}_n=\sum_{j\in\mathcal{B}^{i}_n}A^{ij}_{n-1}.
\end{equation}
Now choose $b^{i}_n$ elements of $\mathcal{G}^{i}_n$ at random
and suppose that $r^{ik}_n$ is the number of times node $k$ is
chosen. The adjacency matrix at the next time step is
\begin{equation}\label{eq:rewiring}
A^{ik}_{n}=\left\{
\begin{tabular}{ll}
      $0$, & $k\in\mathcal{B}^{i}_n  \cup\{i\}$ \\
      $A^{ik}_{n-1}+r^{ik}_n$, & $k\in\mathcal{G}^{i}_n$
     \end{tabular}
     \right..
\end{equation}
It is worth noting that $\mathcal{G}^{i}_n$ contains two sets of nodes: those that were neighbours of $i$ at time
$n-1$ and which were not bad according to the criterion (\ref{eq:criterionbad}) at time $n-1$, and those that were
not neighbours of $i$ at time $n-1$. This means that at each time step with $\mathcal{B}^{i}_n$ non-empty, connections from outside the set of previous
neighbours becomes possible, and also that there is no memory of whether a node has been bad in the past.

In all the cases described here the initial connectivity is
the symmetric all-to-all connectivity where each node in the network is
connected to all the possible $m=N-1$ neighbours and $A_0^{ii}=0$.

\section{Numerical Results}\label{sec:results}

\subsection{Asymptotic network topology}

The first main observation is that, contrary to other models of
adaptive networks of chaotic maps \cite{Ito2001, Ito2003}, in
this model the network reaches a frozen state where the
rewiring stops for all values of $\varepsilon\in[0,1]$. The
existence of the frozen state is partly explained by the
rewiring mechanism chosen (see section \ref{sect:freeze} for further explanation and mathematical proof). If, for instance, a node $i$
receives all its incoming links from one single neighbour $k$
at some iteration $n^{\prime}$, then $\langle
D\rangle^{i}_{n}=D^{ik}_{n}$ and $\mathcal{B}^{i}_n=\emptyset$
for all $n>n^{\prime}$. Therefore $i$ will remain locked to
this neighbour and there will be no further change to this part of the network topology.

\begin{figure}[h]
\centerline{
\includegraphics[width=8cm]{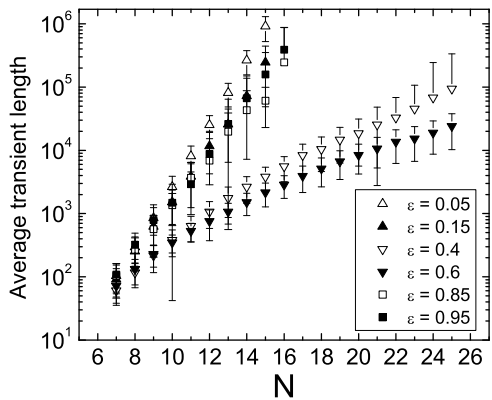}
} \caption{Average transient length to the frozen state as a function of the system
size $N$ for different values of the coupling constant
$\varepsilon$. The average is calculated over 500 realizations
of the system. The frozen state is identified when the network
topology remains constant for $10^{4}$ iterations.}
\label{fig:one}
\end{figure}

\begin{figure}[h]
\centerline{
\includegraphics[width=8cm]{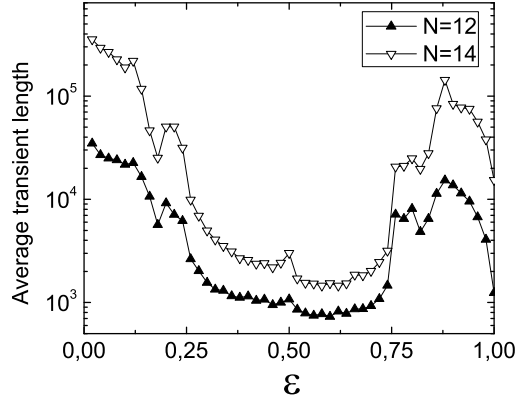}
} \caption{Average transient length to the frozen state as a function of the
coupling constant $\varepsilon$ for different values of the
system size $N$. The average is calculated over 500
realizations of the system. The frozen state is identified when
the network topology remains constant for $10^{4}$ iterations.}
\label{fig:two}
\end{figure}
The duration of the transient to the frozen state appears to increase
exponentially with the system size $N$ (Fig.~\ref{fig:one}) and
depends on the value of $\varepsilon$ (Fig.~\ref{fig:two}). The
dependence of the transient length on the coupling constant is a
sign of the influence of the dynamics in the rewiring and freezing
processes. The exponential increase of the transient time with the
system size is similar to that described in
\cite{kaneko1990supertransients} for the case of a coupled map
lattice with diffusive coupling although the definition of the
transient is different. In the lattice case the topology is fixed
and the transient is defined as the time it takes to reach a certain
attractor.

In both Fig.~\ref{fig:one} and Fig.~\ref{fig:two} there seems to be a marked difference between parameters
$\varepsilon$ in the interval $[0.25,0.75]$ and parameters outside this interval. The transient
times appear significantly shorter for parameters inside this central interval, and as we shall see
(although this is, of course, not an explanation) the dynamics of the nodes for the stationary network
is different in these two cases too.

\begin{figure}[h]
\centerline{
\includegraphics[width=7cm]{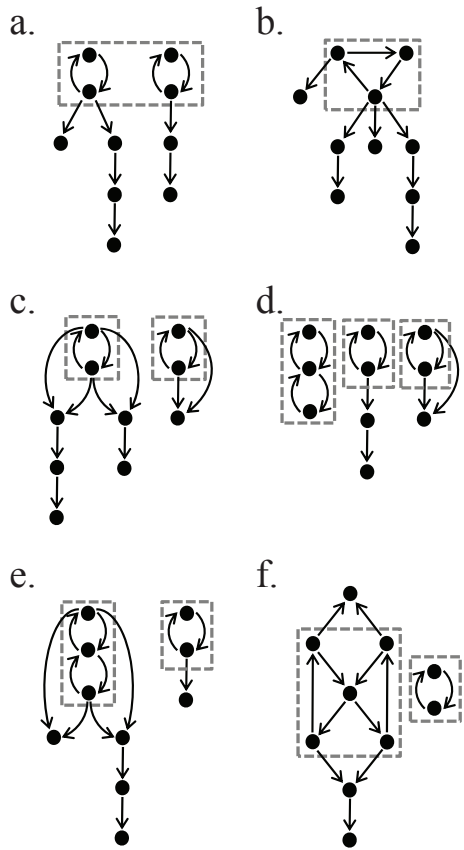}
} \caption{Examples of the final network topology ($N=10$). The
values of the coupling constant are: $\varepsilon=0.1$ in
\textbf{a.} and \textbf{b.}; $\varepsilon=0.4$ in \textbf{c.};
$\varepsilon=0.6$ in \textbf{d.}; $\varepsilon=0.8$ in
\textbf{e.}; $\varepsilon=0.95$ in \textbf{f.}}
\label{fig:three}
\end{figure}

In Fig.~\ref{fig:three} we show six examples of final
topologies of a network of $N=10$ nodes for different values of
$\varepsilon$. The most clear feature of these network examples
is the strong hierarchical structure. This model does not allow
a tree structure as a final topology since all nodes have input
links by definition and therefore the network will have at least one cycle.
However, the structure is very close to the hierarchy of a tree
structure if we consider strongly connected components of the
network as \emph{roots}. (We say a set of nodes is
\emph{strongly connected} if there is a path in the graph
following the directed edges or arrows between any two nodes.)
Inspired by the definitions of `trophic level' and `trophic
height' introduced in \cite{quince2005topological} for the
study of food webs, we can define the `level' of a node as the
minimum (directed) path length from the root to the node and
the `height' of a node as the average distance over all
possible directed paths from the root to the node. We say a
network is strongly hierarchical if level and height coincide
for all the nodes in the network. We can see that following
this definition all the topologies shown in
Fig.~\ref{fig:three} are strongly hierarchical.

The observation of these topologies also allows us to deduce
some dynamical properties of the network. In examples
\textbf{a.} and \textbf{b.} ($\varepsilon=0.1$) all nodes are
locked to one single neighbour. In the remaining
the examples there are nodes with inputs coming
from two different neighbours. As we shall see in the following
sections, this is due to synchronization phenomena in the
strongly connected components. When a node $i$ has only two
neighbours $j,k$ and these are synchronized
($x^j_n=x^k_n$, for all $n$), then $D^{ij}_n=D^{ik}_n=\langle
D\rangle^{i}_{n}$ for all $n$ and the node remains locked to
its neighbourhood. Mathematically this is highly unlikely since usually the orbits synchronize
only eventually and are therefore never exactly the same. However, we find such states due to
machine-precision effects in the numerical computations. In section~\ref{sect:freeze}, where we prove the freezing of the network,
a extra rule for the rewiring mechanism is added to avoid such situations.

\begin{figure}[h]
\centerline{
\includegraphics[width=8cm]{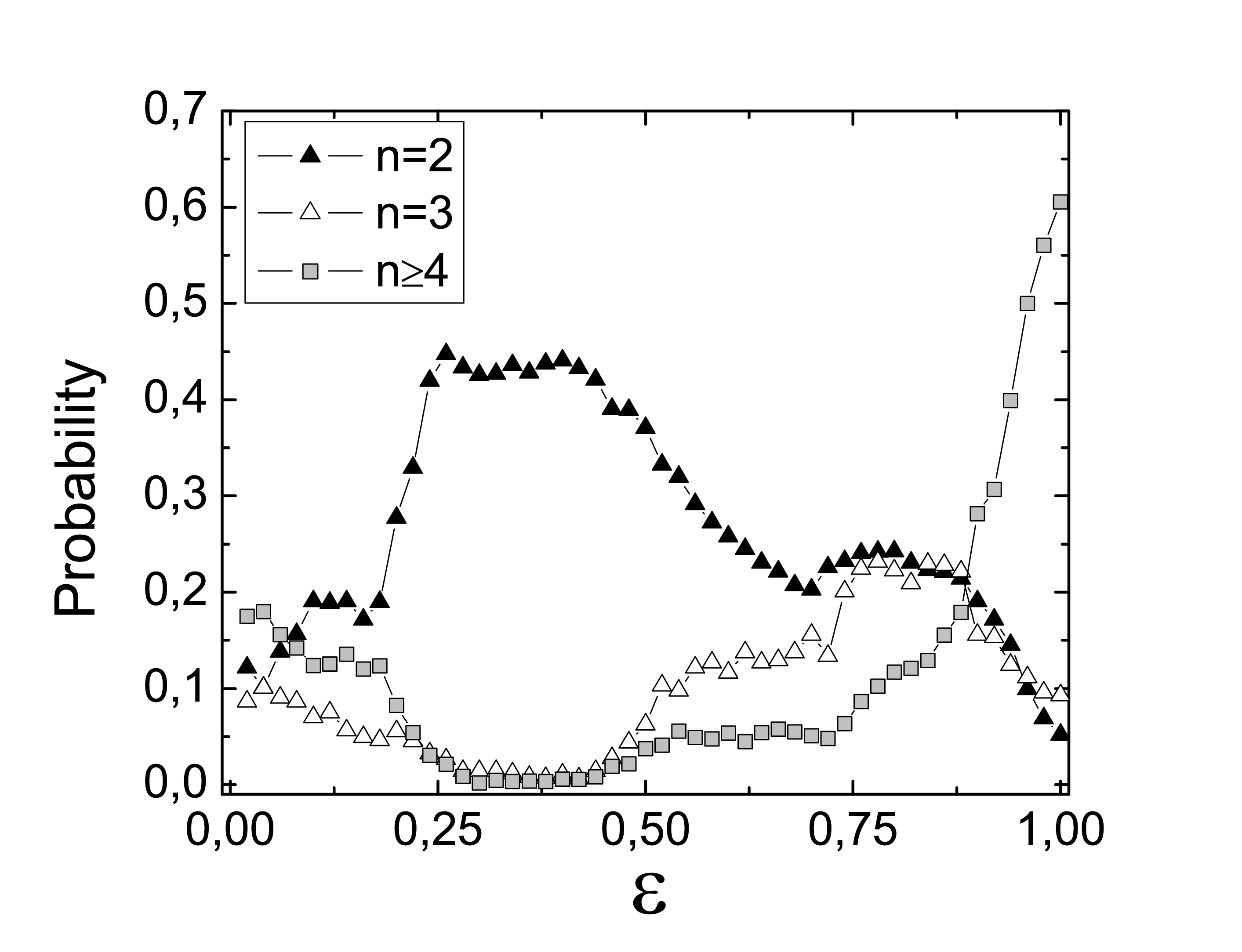}
} \caption{Probability of a node being in a strongly connected
component of size $n$ calculated over 500 initial conditions for each value of $\varepsilon$.} \label{fig:CompProb}
\end{figure}

In Fig.~\ref{fig:CompProb} we show the probability that a node
belongs to a strongly connected component of a certain size as a function of $\varepsilon$. As
could be already appreciated in Fig.~\ref{fig:three}, the most common
strongly connected components are pairs and triplets except in the
region of large $\varepsilon$ ($\varepsilon>0.9$) where bigger
strongly connected components are possible. This and the rest of the
variations of the probabilities with $\varepsilon$ can be better
understood by studying the synchronization dynamics. Therefore, we will come back to
this figure in the next section.

\subsection{Dynamics}

\begin{figure}[h]
\centerline{
\includegraphics[width=8cm]{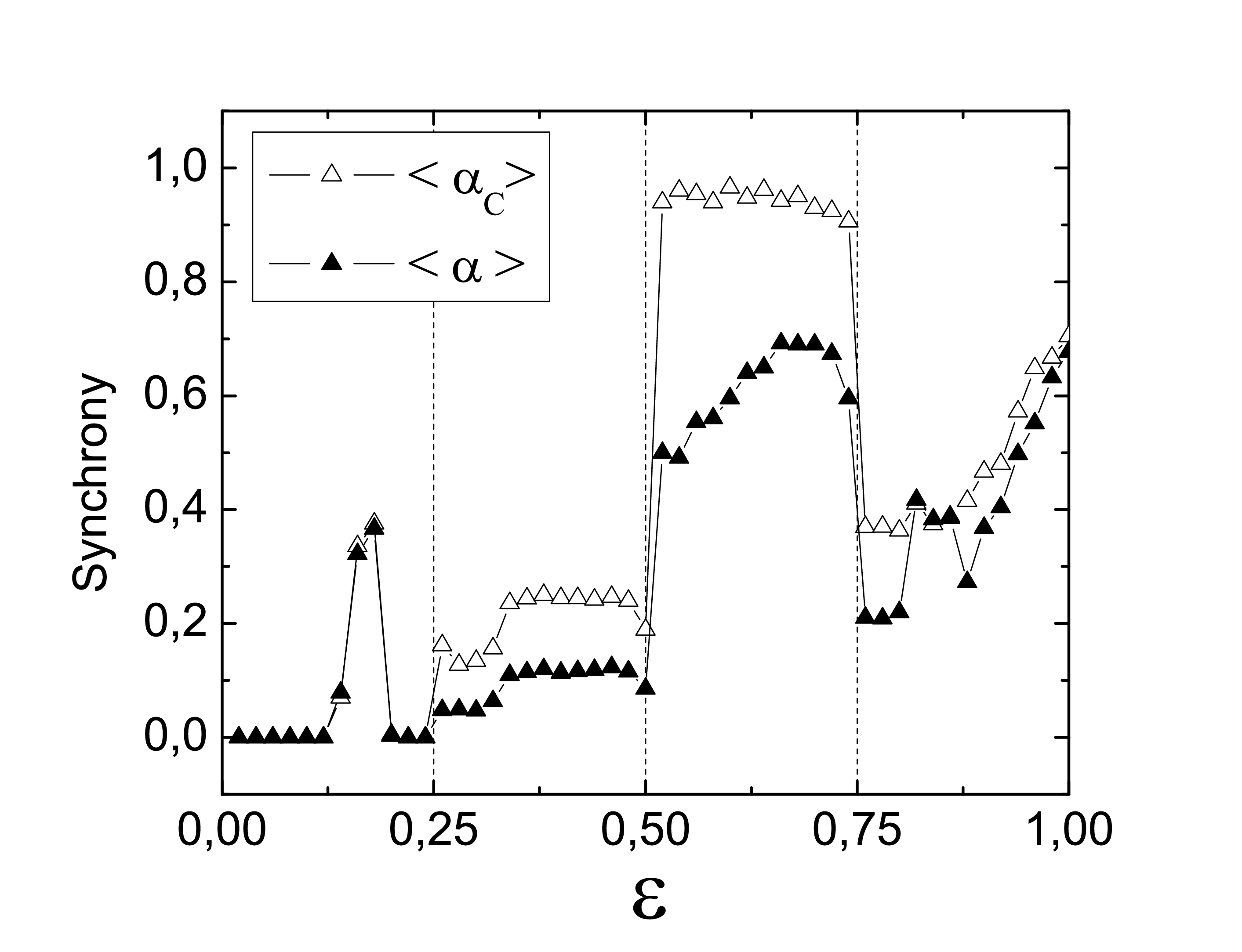}
} \caption{Averages over 500 initial conditions of the
synchrony measures $\alpha$ and $\alpha_C$ as a function of the
coupling parameter $\varepsilon$. The frozen state is
identified when the network topology remains constant for
$10^{4}$ iterations. Parameters: $N=10$, $\tau=10$,
$n^{\prime}=10^{4}-\tau$.} \label{fig:synchrony}
\end{figure}

We study now the dynamics of the nodes in the final network as
a function of the coupling constant. Our main interest is to
see if any nodes in the network synchronize and how they do it.

To measure the synchronization of nodes we find it useful to
define the matrix
\begin{equation}
\beta^{ij}=\theta(\frac{1}{\tau}\sum_{n=n^{\prime}}^{n^{\prime}+\tau}|x^i_n-x^j_n|-\delta)
\end{equation}
where $\theta(x)$ is the Heaviside step function, $n^{\prime}$ is a long transient that we allow in order
to be sure the network has frozen and the dynamics have
stabilized and $\delta$ is a small quantity that we introduce to properly detect eventually synchronous dynamics (numerically, machine-precision effects do the work).  The element $\beta^{ij}$ is equal to zero if the
trajectories of nodes $i$ and $j$ are fully synchronized
($x^i_n=x^j_n$) during $\tau$ iterations after the transient
and is equal to one otherwise.

We can now define a measure of full synchronization of the
network as
\begin{equation}
\alpha=1-\frac{1}{N(N-1)}\sum_{\substack{i,j\\i\neq
j}}\beta^{ij}.
\end{equation}
This measures the percentage of synchronized pairs of nodes
(connected or not) over the total number of pairs. If
$\alpha=1$ all nodes in the network are synchronized in the
same trajectory while if $\alpha=0$ no two nodes in the network
are synchronized.

Since our network can split in several disconnected
components and each connected component could be fully
synchronized in a different trajectory, we introduce a second quantity to take
this into account and measure the synchronization only between
pairs of connected nodes. We can thus define the connected
component synchronization as
\begin{equation}
\alpha_C=1-\frac{1}{|C|}\sum_{\substack{(i,j)\in C}}\beta^{ij}
\end{equation}
where $C$ is the set of pairs of connected nodes and $|C|$ is
the cardinality of this set.

\begin{figure}[h]
\centerline{
\includegraphics[width=6cm]{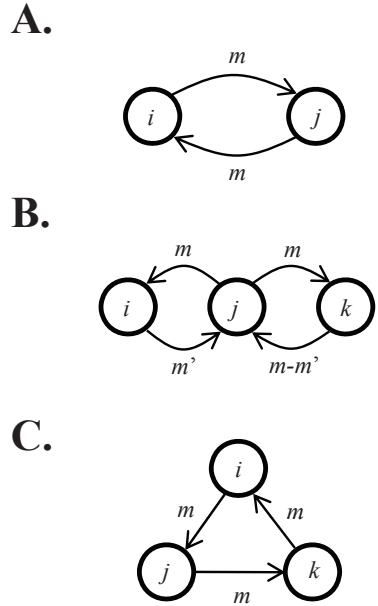}
} \caption{\textbf{A.} Completely connected pair (2-cycle).
\textbf{B.} Triplet with transposition symmetry. \textbf{C.}
3-cycle.} \label{fig:four}
\end{figure}

We can see the values of $\alpha$ and $\alpha_C$ as a function
of $\varepsilon$ in Fig.~\ref{fig:synchrony}. To explain the
different regimes in this figure it is very useful to study
first the dynamics of the most common small strongly connected
components such as the completely connected pair, the triplet
with transposition symmetry and the 3-cycle shown in
Fig.~\ref{fig:four}. Since these act as \emph{roots} from which
the rest of the network takes their inputs, the dynamics of
these components is what determines the behaviour of the rest
of the nodes. In appendix \ref{app:dynamics} we detail the
calculations. Here we will only report the results that are of
interest for the discussion. The Lyapunov exponent of the
fully-chaotic logistic map ($r=4$) is $\lambda=\ln 2$.
Substituting this in (\ref{eq:pair}) we find that the
synchronous chaotic state of the completely connected pair is
stable in the interval $0.25<\varepsilon<0.75$. Similarly, the
triplet with transposition symmetry (\ref{eq:triplet}) has a stable synchronous
state if $0.5<\varepsilon<0.75$. On the other hand, for the fully chaotic logistic map ($r=4$) used here
the 3-cycle (\ref{eq:cycle}) has no stable synchronous state. Another important
fact is that a node locked to a synchronized set of nodes (all following
an orbit of the uncoupled logistic map), as in (\ref{eq:unidirectional}), will synchronize to them
if $\varepsilon>0.5$.  This results makes the
interpretation of Fig.~\ref{fig:synchrony} much more straightforward. The change of
regime at $\varepsilon=0.25$ is explained by the strongly
connected pairs becoming synchronized. Also, in Fig.~\ref{fig:CompProb} we can see that the probability of finding pairs in the final network greatly increases. At $\varepsilon=0.5$ the
synchronized state of the triplet with transposition symmetry
becomes stable. This is likely to be the cause of the increase, at $\varepsilon=0.5$, of the probability of being in a strongly connected component of size $n=3$. Moreover, a node locked to a synchronized pair
or triplet will become synchronized with it and due to the
hierarchical structure of the networks, this opens the
possibility for the whole network to synchronize in the same
orbit. When $\varepsilon=0.75$ the pair and the triplet
synchronized states both lose stability. However, Fig.~\ref{fig:synchrony} shows that
in both $0.12\lesssim\varepsilon\lesssim0.2$ and
$\varepsilon>0.75$ there is a considerable amount
of synchronized nodes in the final networks even though none of the most common strongly connected components
has a stable synchronous state. This is partly caused by the
phenomena of polysynchrony that we explain in detail in the
next section.

\subsection{Polysynchrony}

In most studies of synchronization on networks, if two or more nodes synchronize then they are connected directly in the network, and the synchronized
states form clusters. The term polysynchrony \cite{stewart2003symmetry,golubitsky2004some} was introduced to describe a more general form
of synchronization on networks for which the synchronized states are not necessarily directly connected within the network. Examples, and
further analysis of general conditions for the existence of such states can be found in
\cite{stewart2003symmetry,golubitsky2004some,field2004combinatorial,aguiar2009dynamics, agarwal2010dynamical}.

\begin{figure}[h]
\centerline{
\includegraphics[width=6cm]{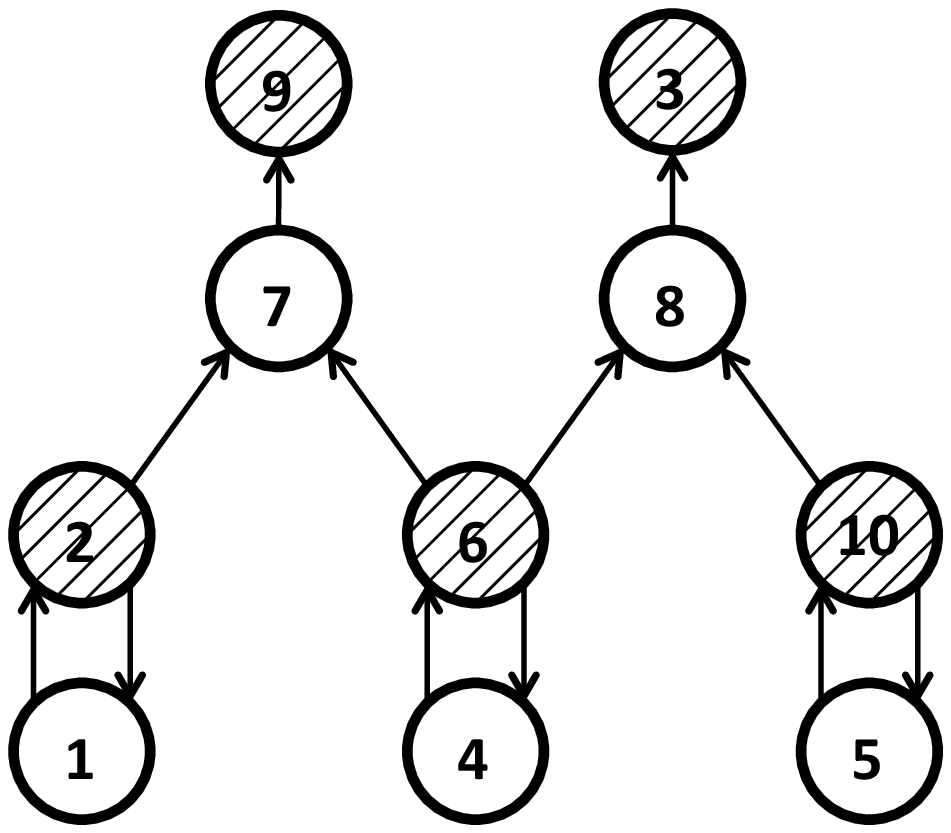}
} \caption{Example of polysynchronous network for
$\varepsilon=0.85$. Nodes filled with the same pattern are synchronous. In this case each synchrony class is attracted to a different fixed point.} \label{fig:exFP}
\end{figure}

\begin{figure}[h]
\centerline{
\includegraphics[width=8cm]{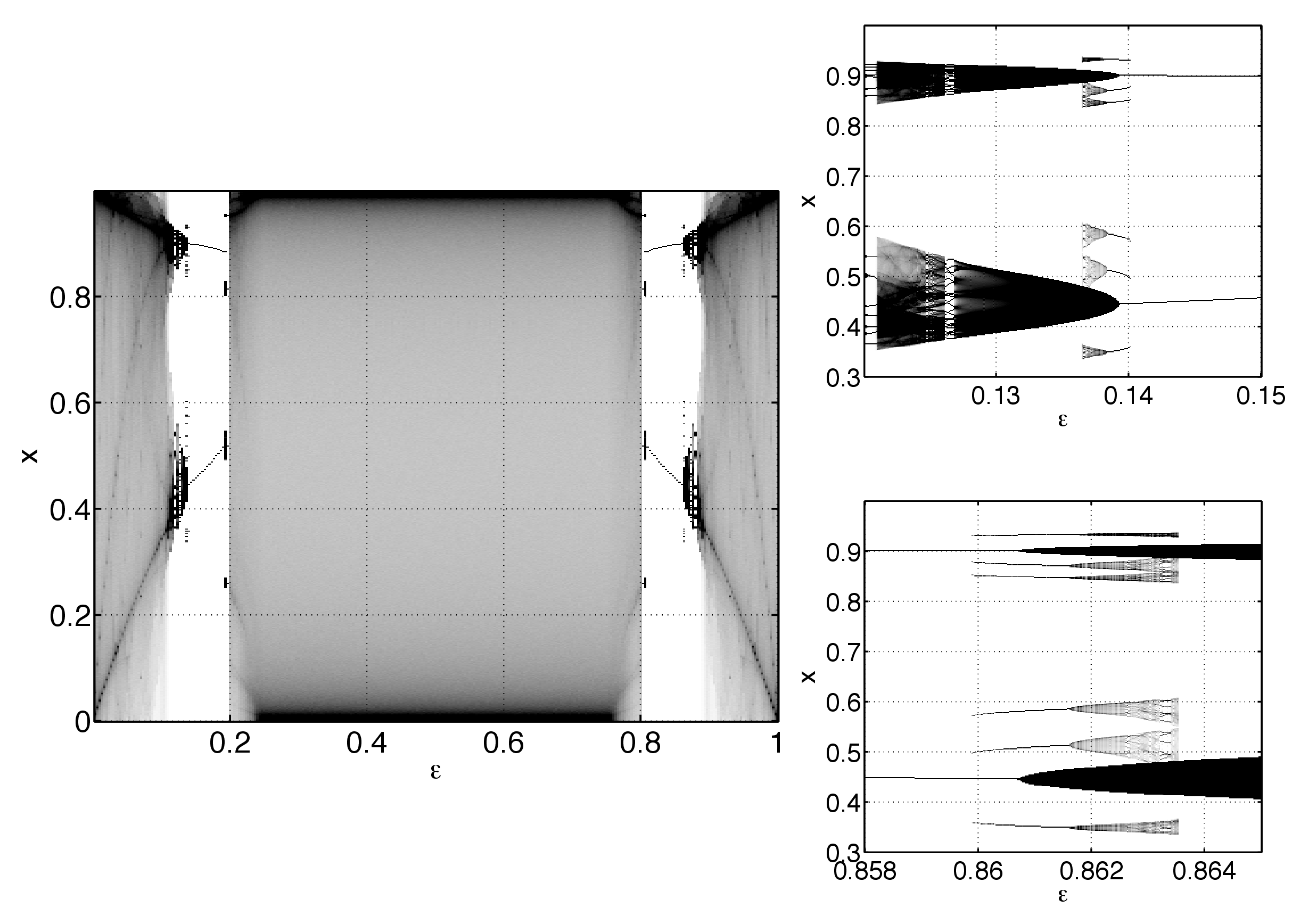}
} \caption{Bifurcation diagram as a function of $\varepsilon$
for the system of two coupled fully-chaotic logistic maps (\ref{eq:pair}).}
\label{fig:coupledLog}
\end{figure}

We illustrate now the phenomenon of polysynchrony in our model with several examples from the simplest case of fixed point dynamics to more involved examples of quasiperiodic and chaotic polysynchronous dynamics.  In Fig.~\ref{fig:exFP}, for $\varepsilon=0.85$, we find that each synchrony class has a fixed point as the final attractor. These fixed points correspond to the fixed point dynamics of the completely connected pair since the root of the network in this example is composed of three completely connected pairs.  In fact, for most of the examples of polysynchrony provided the quotient system of the network, obtained
 by identifying synchronized nodes, reduces to a completely connected pair \cite{bg2011polyletter}. Therefore, the available dynamics are those of the completely connected pair (see Fig.~\ref{fig:coupledLog}).

\begin{figure}[h]
\centerline{
\includegraphics[width=8.5cm]{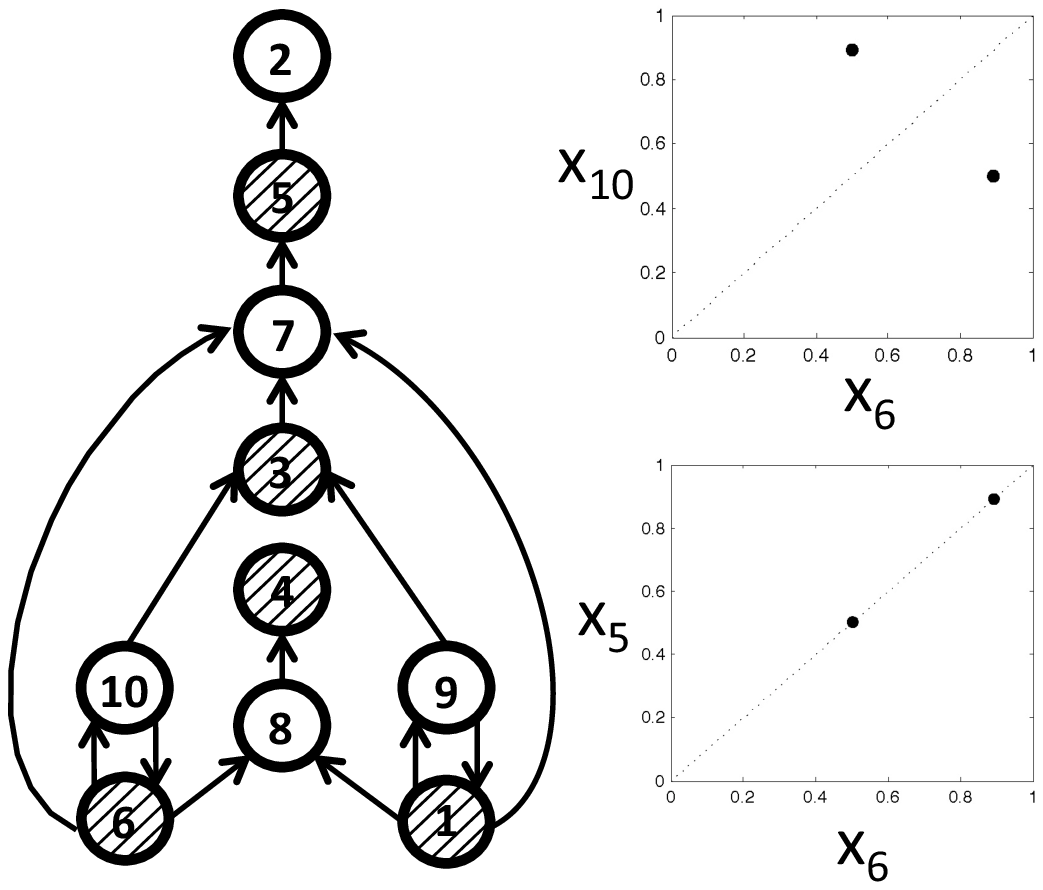}
} \caption{Example of polysynchronous network for
$\varepsilon=0.18$. Nodes filled with the same pattern are synchronous. The dynamics of the synchrony classes is periodic with period 2. Nodes in different classes oscillate in antiphase.} \label{fig:exP2}
\end{figure}

For $\varepsilon=0.18$, in Fig.~\ref{fig:exP2} we find a network with polysynchronous period-2 dynamics. The dynamics is divided into two synchrony classes following the same period-2 orbit in antiphase. Since for a
given $\varepsilon$ the dynamics of all pairs is the same, the
nodes of a pair are synchronous with the corresponding nodes of
the other pair.

\begin{figure}[h]
\centerline{
\includegraphics[angle=-90,width=8cm]{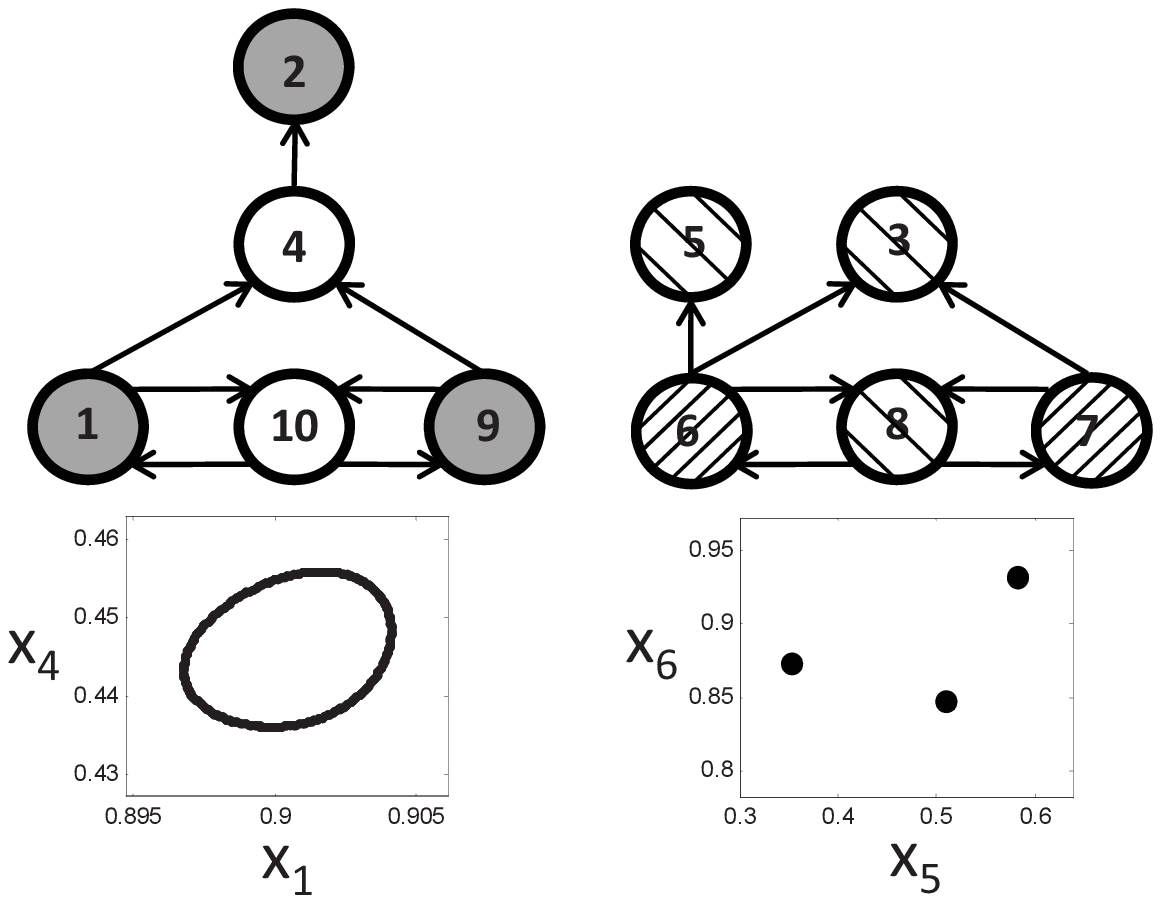}
} \caption{Example of polysynchronous network for $\varepsilon=0.861$. Nodes filled with the same pattern are synchronous. The nodes in the left cluster follow quasiperiodic orbits while the nodes in the right cluster follow period-3 orbits.} \label{fig:example1}
\end{figure}

In Fig.~\ref{fig:example1} (for $\varepsilon=0.861$), the network has divided
into two separate clusters. In one of them the dynamics of the
nodes is quasiperiodic while in the other it is periodic with
period 3. Both clusters have a triplet with transposition
symmetry as a root. As in the previous examples no two synchronized nodes are connected and all nodes with equivalent inputs are
synchronized. Although the two clusters have different dynamics, their quotient systems are completely connected pairs and therefore, both the period-3 and the quasiperiodic orbit are attractors of the completely connected pair (see Fig.~\ref{fig:coupledLog}) when $\varepsilon=0.861$.

\begin{figure}[h]
\centerline{
\includegraphics[angle=-90,width=7cm]{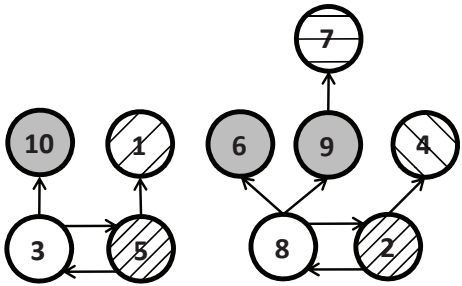}
} \caption{Example of polysynchronous network for
$\varepsilon=0.14$. Nodes filled with the same pattern are synchronous. In this case the nodes in the completely connected pair follow period-2 orbits as in Fig.~\ref{fig:exP2} and the rest of the nodes follow different period-4 orbits.}
\label{fig:example2}
\end{figure}

Fig.~\ref{fig:example2} shows a slightly different example of
polysynchrony. As in Fig.~\ref{fig:example1}, the network has
split into two clusters. The roots of the clusters are completely
connected pairs. The dynamics of the nodes in the pairs is
periodic with period 2 as in Fig.~\ref{fig:exP2}. The two nodes inside the pair follow
the same orbit but they are out of phase. The dynamics of the rest of the nodes in the
network is periodic with period 4. Interestingly, in this
example we can see how nodes with the same input (such as nodes
$1$ and $4$) do not necessarily synchronize.

\begin{figure}[h]
\centerline{
\includegraphics[width=8.5cm]{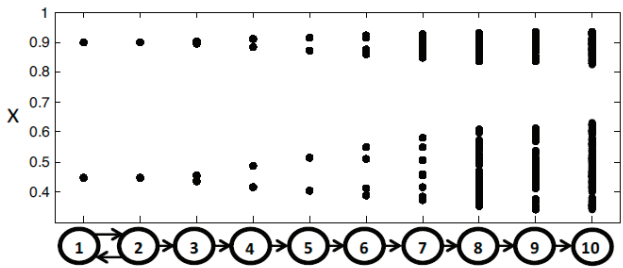}
} \caption{Example of network similar to an open flow system with a completely connected pair at one end that forces the system with a period-2 orbit. We can see how the dynamics of the nodes follow a spatial route-to-chaos along the chain. In this example
$\varepsilon=0.14$.}
\label{fig:openflow}
\end{figure}

This particular case is more involved because we find here an instance of spatial route to chaos in an open flow similar to that described in \cite{kaneko1985spatial, willeboordse1995pattern, rudzick1996unidirectionally, yamaguchi1997mechanism}. In Fig.~\ref{fig:openflow} we observe the prototypical open flow system, consisting of a chain of unidirectionally coupled maps. In this case the system is closed on one side by a completely connected pair. The period-2 orbit of the pair for $\varepsilon=0.14$ is fed into the chain as a fixed boundary condition and we observe a spatial period-doubling bifurcation. What we observe in Fig.~\ref{fig:example2} is merely the beginning of this route-to-chaos.

\begin{figure}[h]
\centerline{
\includegraphics[width=7cm]{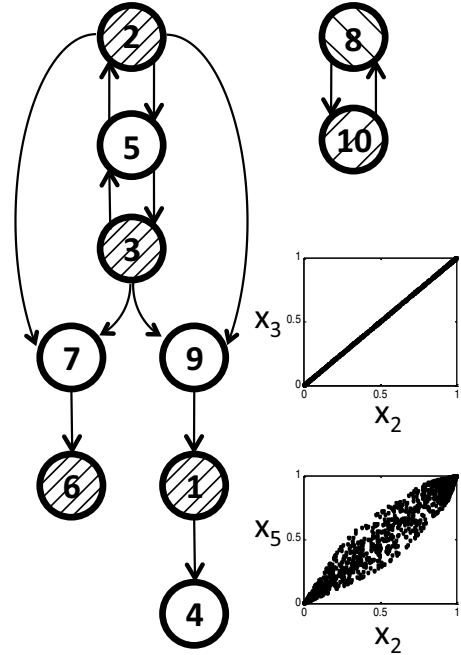}
} \caption{Example of polysynchronous network for
$\varepsilon=0.78$. Nodes filled with the same pattern are synchronous. The dynamics of both synchrony classes is chaotic in this case.}
\label{fig:example3}
\end{figure}

Fig.~\ref{fig:example3} shows an example of polysynchronous
network for $\varepsilon=0.78$ where the dynamics of the nodes is chaotic. It is important to
note that the synchronized trajectories do not correspond to
trajectories of the uncoupled logistic map. If this were the case, nodes 7 and 9 would synchronize to nodes 2 and 3 since $\varepsilon>0.5$ (see appendix \ref{appsub:uni}).

\begin{figure}[h]
\centerline{\includegraphics[width=8cm]{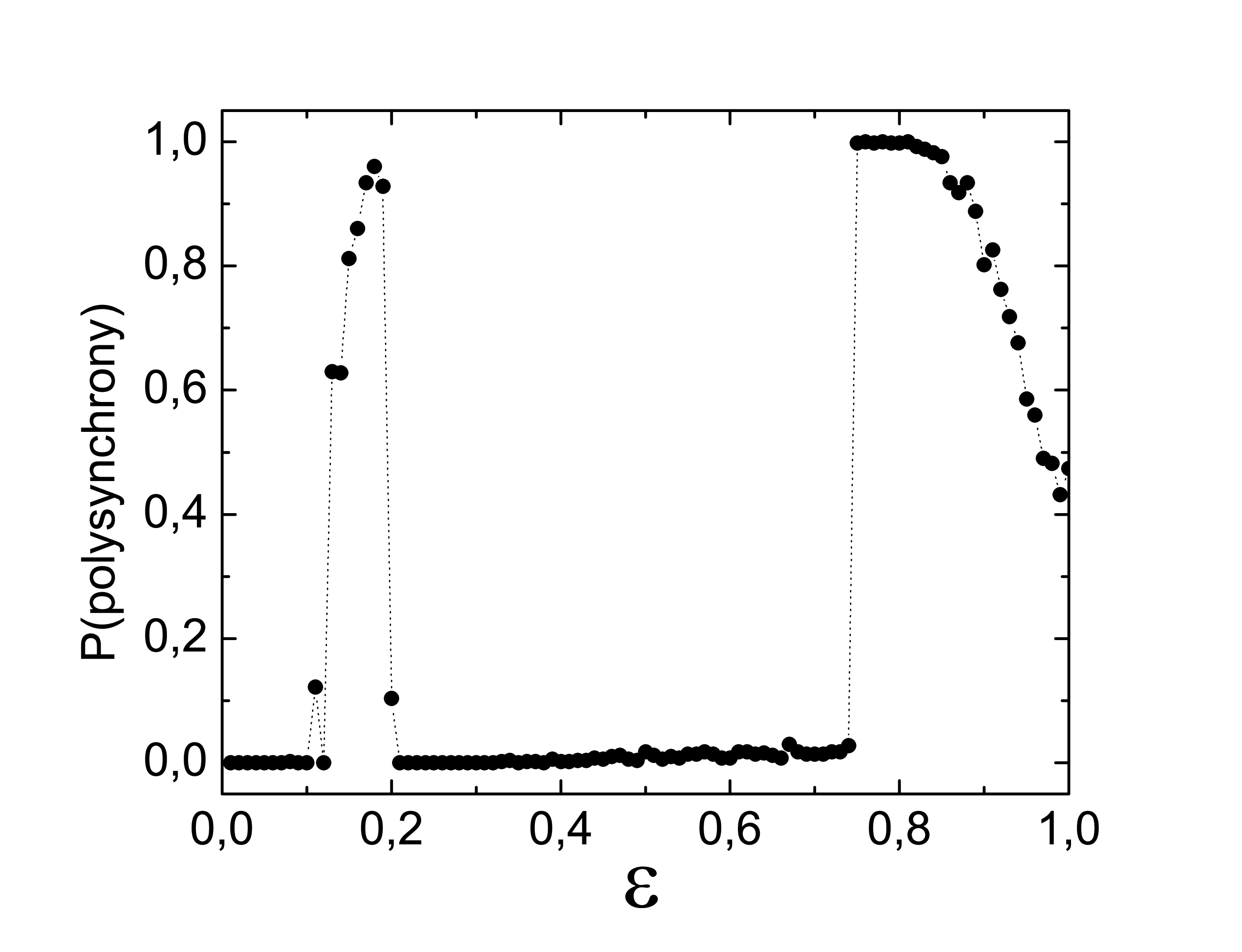}
} \caption{Percentage of final network topologies of size
$N=10$ showing polysynchronous patterns calculated over 500
realizations of the system as a function of $\varepsilon$. (Reprinted from \cite{bg2011polyletter} with permission.)}
\label{fig:ProbPoly}
\end{figure}

Finally, in Fig.~\ref{fig:ProbPoly} we show the probability of finding
polysynchrony in the network as a function of the coupling
constant $\varepsilon$. As we have already detected in the study of synchrony in Fig.~\ref{fig:synchrony}, there are two intervals of coupling strength values for which polysynchrony is possible.

The regions of polysynchrony in Fig.~\ref{fig:ProbPoly} can be better understood by studying the stability of the different polysynchronous dynamics.  In Appendix \ref{app:poly} we study the stability of the simplest (and more common) polysynchronous states in the triplet with transposition symmetry which is one of the smallest structures that can show polysynchrony. As explained in the appendix, the obtained results are independent of the value of $m'$ (see Fig.~\ref{fig:four}\textbf{B.}) and therefore, are also valid for the case of a completely connected pair with a third node unidirectionally coupled to one of the nodes in the pair ($m'=m$ or $m'=0$).  The polysynchrony in the interval $0.12\lesssim\varepsilon\lesssim0.20$ is mainly explained by the stability of the period-2 polysynchronous state. This state is stable in the range $0.140375\lesssim\varepsilon\lesssim0.193814$. Although this state occupies most of the interval, quasiperiodic polysynchronous states and periodic polysynchronous states of higher period can be numerically witnessed (Fig. \ref{fig:coupledLog}). As we have also determined analytically, chaotic polysynchrony with coupled fully-chaotic logistic maps ($\lambda=\ln 2$) is only possible for $\varepsilon>0.75$. This fact, together with the stability of the fixed-point polysynchronous state in the range $0.806186\lesssim\varepsilon\lesssim0.86$ accounts for most of the polysynchrony found for $\varepsilon>0.75$. However, as before, other periodic and quasiperiodic polysynchronous states can be found in that region.

\section{Analysis of the frozen state}
\label{sect:freeze}

As noted earlier in the discussion of Fig. \ref{fig:three} some numerical simulations lead to stationary (frozen) networks in which one or more nodes has more than one
input. Mathematically this can only happen if there is complete synchronization between the input nodes, not simply that the values approach
each other. This is highly unlikely, and the fact that we find these configurations in examples reflects the speed of convergence to
synchronization and the finite precision of the simulations. To avoid this possibility, and to make our proofs simpler, we use a slightly modified
rewiring rule in this section. Equations (\ref{eq:criterionbad})-(\ref{eq:rewiring}) specify how to change the network structure if there exists
$j$ with $D_n^{ij}>\langle D\rangle_n^i$. We now add an additional rule to resolve the ambiguity that arises if node $i$ has more than one neighbour and $D_n^{ij}=\langle D\rangle_n^i$ for all of these neighbours. The additional
rule, which we will refer to as rule (R), states:

\medskip
\textbf{(R)}\emph{If ${\cal B}_{n}^i=\emptyset$ and $i$ has more than one neighbour then choose a neighbour $k_i$ at random and set}
\begin{equation}
A_n^{ij}=\left\{\begin{array}{ll} m & {\rm if}~j=k_i\\ 0 & {\rm if}~j\ne k_i\end{array}\right.
\end{equation}

Thus if all the neighbours are `good', which would have led to no rewiring in the previous rule, we choose one of these at random and rewire
all inputs to node $i$ from this choice. Of course, once this is done there can be no further rewiring (as the node $k$ is compared only to itself)
and so the connection to node $i$ is from a single node $k_i$. In terms of the polysynchronous states observed numerically, this rule would lead to further evolution in the network topology of the polysynchronous networks shown in the previous section but polysynchrony would still occur (although in simpler networks).

Note that since the dynamics has a stochastic component it is not surprising that the freezing theorem
will also be probabilistic: we will prove that the probability that the network has not frozen by time $n$,
i.e. the dynamics of the network (but not necessarily the dynamics on the nodes)
is stationary from time $n$ onwards, tends to zero as $n$ tends to infinity.

Before writing down the detailed calculations we will describe the strategy of the proof. We begin by considering a slightly modified system; one that is realized with non-zero probability in the
dynamics described above. The finite probability system analyzed here is a subsystem of the general case in which at each time step, every connection that is rewired is rewired to the good node
to which it already has the most connections (or one of these at random if there are two or more such nodes). At each time step, either a node has only one neighbour, and there can be no rewiring, or the number of connections
to the most connected node increases by at least one. Since each node has $m$ inputs, there will be one neighbour within $m$ time steps. At each time step this happens in
the real system with a probability that is bounded below by a fixed non-zero $p$. Hence for any finite $T$ there is a finite probability
(greater than $p^T$) that this revised rule will be used at each of the next $T$ time steps and hence, as $n$ goes to infinity, the probability of freezing goes to one. Returning to the original system this modified system occurs for $m$ times steps with a finite probability, and hence the probability that this modified
rule is applied is non-zero and the probability that the original system does not freeze tends to zero as time goes to infinity. We will now provide
the details.

The modified system is specified as follows. At each time step $n$ the mean distance $\langle D\rangle_n^i$ is calculated which determines the good set
and bad set, ${\cal G}_n^i$ and ${\cal B}_n^i$ for each $i\in\{1,\dots ,N\}$ as explained in section \ref{sec:model}, equations (\ref{eq:distance})-(\ref{eq:criteriongood}). Now, the rewiring condition (\ref{eq:criterionbad}) implies that at least one of the nodes that is `good' for a given node at time $n-1$ is also `good' for that node at time $n$, so it is always possible to choose a node $k(i,n)\in {\cal G}_n^i$
such that
\[A_{n-1}^{ik(i,n)}=\max_j A_{n-1}^{ij}
\]
where the maximum is over $j\in {\cal G}_{n-1}^i\cap {\cal G}_n^i$, and if the maximum is attained by more than one node, one of these is chosen at random. Then
if $b_n^i$ is the valency of the bad nodes as defined in (\ref{eq:bvalency}) then
\begin{equation}\label{eq:modrule}
A_n^{ij}=\left\{\begin{array}{ll} 0 & {\rm if}~j\in {\cal B}_n^i\cup\{i\}\\ A_{n-1}^{ij} & {\rm if}~j\in {\cal G}_n^i\backslash\{k(i,n)\} \\A_{n-1}^{ij}+b_n^i & {\rm if}~j =k(i,n) .\end{array} \right.
\end{equation}
If $b_n^i\ne 0$ then by definition $A_{n}^{ik(i,n)}>A_{n-1}^{ik(i,n-1)}$ and so (since they are bounded by $m$) after at most $m$ iterations for each $i$ there exists $k$, and $r\le m$ such that $A_r^{ik}=m$, $b_r^i=0$ and $A_s^{ik}=m$ for all $s>r$. In other words the network has frozen. 

This rule \emph{could} be the outcome of the original rules when the bad set is non-empty if all but one of the numbers $r_n^{ij}$ were zero and
so the remaining $r_n^{ij}$ equals $b_n^i$ and this final connection is to a particular chosen node (that with the largest current connectivity to $i$).
If there are $s$ bad nodes and $N-s$ good nodes, then for a given $i$ the probability of picking the `right' good node is $1/(N-s)$ and so the probability of
rewiring all the bad nodes to this node is $1/(N-s)^{b_n^i}$. Now, $b_n^i\le m$ and $N-s\le N$ so the probability of making this choice is greater
than $(1/N)^m$. This is true for each of the $N$ nodes labelled by $i$ and so the probability of the original system behaving in this way in one time step is
greater than $(1/N)^{mN}$.

Now consider using the modified rule (\ref{eq:modrule}) together with the additional rule (R). Then at each time step either
the number of connections of the most connected node to $i$ increases by at least one, or there is only one node connected to $i$ and so there
can be no further changes to the connections to $i$. Since there are a total of $m$ connections to each node,
this latter state must be achieved within $m$ time steps
of this modified system, after which it is frozen (and it is frozen whichever rules are used after this stage).

Now, the probability of applying this modified rule to the original system for $m$ consecutive time steps is just  $(1/N)^{m^2N}$, so if we return
to the original system with our additional rule (R), time can be divided up into segments of length $m$, and so in time $rm$ there are $r$
independent opportunities to use the modified rule that leads to freezing, each with probability greater than $(1/N)^{m^2N}$,
where the extra factor of $m$ in the exponent reflects the fact that the modified rule is applied at most $m$ times to obtain the frozen
state. So the probability that
the system does not freeze in time $rm$ is less than
\begin{equation}
\left(1-\frac{1}{N^{m^2N}}\right)^r
\end{equation}
which obviously tends to zero as $r\to \infty$, completing the proof.

The estimate of the probability could be improved considerably, for example by considering overlapping time intervals, but we are only interested in whether the probability of this not happening tends to zero, and for this the argument above suffices and has the virtue of simplicity.

Note that each node of the frozen network topology has precisely one neighbour. From this it is easy to show that each connected component
of the network has one and only one strongly connected component (a cycle) and then trees based on the elements of the cycle. This means that the
network eventually has precisely the hierarchical structure of \cite{quince2005topological} when the cycle is considered as the root of the network.

\section{Discussion}\label{sec:discussion}

In this paper we have studied the dynamics of a simple adaptive network model as a function of the coupling parameter. We have rigorously proved that the network reaches a frozen state where the rewiring stops. We have shown that the final topologies are usually hierarchical and that polysynchronous dynamics appear in the frozen networks for certain parameter values. The hierarchical structure of the networks facilitates the appearance of polysynchrony as a stable attractor of the dynamics by making it easier to establish a balanced equivalence relation on the nodes. The stability study of different polysynchronous states explains the concrete coupling parameter ranges for which polysynchrony can be observed.

Unlike many adaptive network studies, the system described here evolves from a totally connected initial state to a much more
constrained final topology. This simplifying structure could be relevant to the formation of functional groups in social
interactions of biological systems. The dynamics on the network also has rich features; so far as we are aware this is the first network
which can evolve naturally to a polysynchronous state. Such states could describe a form of functional evolution where a uniform population
separates into different functional units described by different synchrony classes. The novel feature of polysynchrony is that
these groups do not have to separate spatially as in the standard clusters, which are directly connected within the network.
From this point of view, a fast time adaptive network of the type described here (to establish differentiated populations amongst a
uniform set of initial nodes) followed by a slow differentiation process to lock in the differences created by
the different synchrony classes could be a model for processes that require a mixed heterogeneous population from an initially
homogenous set. Our models bear some resemblance to models in population dynamics (metapopulations, see \cite{parthasarathy1998synchronisation})
and so this may be another area where polysynchrony might arise.

\appendix

\section{Synchronization dynamics of strongly connected components}\label{app:dynamics}

To study the stability of the synchronized state of the
different strongly connected components we will follow the
approach exposed in \cite{pikovsky2003synchronization}. Some of these results are well documented in the literature and are shown here for the sake of completeness.

In all the cases we study the coupling is linear and can be written,
in general, as
\begin{equation}
x^i_{n+1}=\sum_{j=1}^{N}L^{ij}f(x^{j}_n),
\end{equation}
where $L$ is the coupling linear operator. The synchronous state exists if the operator $L$ has an
eigenvalue $\sigma_1=1$ corresponding to the eigenvector
$\textbf{e}_1=(1, 1, \dots, 1)$. Since our coupling is
dissipative, the rest of the eigenvalues of the coupling
operator are in modulus less than one.  The stability of the
synchronized state is then given by the condition
\begin{equation}
\lambda_{\perp}=\lambda+\ln|\sigma_2|<0,
\end{equation}
where $\lambda_{\perp}$ is the transverse lyapunov exponent,
$\lambda$ is the lyapunov exponent of the uncoupled map and
$\sigma_2$ is the second largest eigenvalue of the coupling
operator.

\subsection{Dynamics of the completely connected pair}

The completely connected pair (Fig.~\ref{fig:four}\textbf{A.}) forms a system of two coupled
logistic maps
\begin{equation}\label{eq:pair}
\left(\begin{array}{c} x^{i}_{n+1}\\
x^{j}_{n+1}\end{array}\right)=\left(\begin{array}{cc} 1-\varepsilon & \varepsilon\\
\varepsilon & 1-\varepsilon\end{array}\right)\left(\begin{array}{c}f(x^{i}_{n})\\
f(x^{j}_{n})
\end{array}\right).
\end{equation}

This system has been thoroughly studied as a model of
population dynamics in \cite{gyllenberg1993does,
lloyd1995coupled, kendall1998spatial}.

In this case the linear operator has eigenvalues $\sigma_1=1$
($\textbf{e}_1=(1,1)$) and $\sigma_2=1-2\varepsilon$
($\textbf{e}_2=(-1,1)$). Thus the stability condition reads
\begin{equation}
\lambda_{\perp}<0\rightarrow\left\{
\begin{tabular}{ll}
      $\lambda+\ln(1-2\varepsilon)<0$, & $\varepsilon<\frac{1}{2}$, \\
      $\lambda+\ln(2\varepsilon-1)<0$, & $\varepsilon>\frac{1}{2}$.
     \end{tabular}
     \right.
\end{equation}
Therefore, the synchronization of the pair is stable when
\begin{equation}
\frac{1-e^{-\lambda}}{2}<\varepsilon<\frac{1+e^{-\lambda}}{2}.
\end{equation}

\subsection{Dynamics of the triplet with transposition
symmetry}

The linear operator in the case of the triplet with transposition symmetry (Fig.~\ref{fig:four}\textbf{B.}) is
\begin{equation}\label{eq:triplet}
L=\left(\begin{array}{ccc}1-\varepsilon & \varepsilon & 0\\
\varepsilon\frac{m^{\prime}}{m}
&1-\varepsilon&\varepsilon\frac{m-m^{\prime}}{m}\\
0  & \varepsilon & 1-\varepsilon
\end{array}\right),
\end{equation}
with eigenvalues
\begin{eqnarray*}
\sigma_1&=&1,\\
\sigma_2&=&1-\varepsilon,\\
\sigma_3&=&1-2\varepsilon,
\end{eqnarray*}
corresponding to the eigenvectors $\textbf{e}_1=(1,1,1)$,
$\textbf{e}_2=(\frac{m^{\prime}-m}{m^{\prime}},0,1)$,
$\textbf{e}_3=(1,-1,1)$.

We should note that which eigenvalue has the second largest
modulus depends on the value of $\varepsilon$ and the stability
condition has to be evaluated for both $\sigma_2$ and
$\sigma_3$. It is an easy calculation to deduce that the
synchronized chaotic state will be stable in the range
\begin{equation}
1-e^{-\lambda}<\varepsilon<\frac{1+e^{-\lambda}}{2}.
\end{equation}

\subsection{Dynamics of the 3-cycle}

The linear operator of the 3-cycle (Fig.~\ref{fig:four}\textbf{C.})reads
\begin{equation}\label{eq:cycle}
L=\left(\begin{array}{ccc}1-\varepsilon & \varepsilon & 0\\0
&1-\varepsilon &\varepsilon\\ \varepsilon & 0 & 1-\varepsilon
\end{array}\right),
\end{equation}
and has eigenvalues
\begin{eqnarray*}
\sigma_1&=&1,\\
\sigma_2&=&\frac{1}{2}(2-3\varepsilon+i\varepsilon\sqrt{3}),\\
\sigma_3&=&\frac{1}{2}(2-3\varepsilon-i\varepsilon\sqrt{3}).
\end{eqnarray*}
Thus, the stability condition of the synchronous state reduces to
\begin{equation}
\lambda+\ln|\sigma_2|<0\rightarrow\lambda+\ln\sqrt{1-3\varepsilon+3\varepsilon^{2}}<0.
\end{equation}
Solving this for $\varepsilon$ provides us with the condition
\begin{equation}
\frac{1}{2}-B< \varepsilon <\frac{1}{2}+B,
\end{equation}
where
\begin{equation}
B=\frac{1}{2\sqrt{3}}e^{-2\lambda}\sqrt{-e^{2\lambda}(e^{2\lambda}-4)}.
\end{equation}
Therefore, the stability region for the synchronous state of the 3-cycle is an interval centered around
$\varepsilon=0.5$ of a width depending on the lyapunov exponent
$\lambda$ of the map. When $\lambda=\ln 2$, $B$ vanishes and
the synchronous state becomes unstable for all $\varepsilon$.

\subsection{Dynamics of the unidirectional coupling}\label{appsub:uni}

Apart from the dynamics of the strongly connected components,
it is necessary to study the case where a node is influenced by
a single neighbour following an orbit of the uncoupled logistic map or,
equivalently, by a fully synchronized neighbourhood. In both
cases the dynamics is given by
\begin{equation}\label{eq:unidirectional}
\left(\begin{array}{c} x_{n+1}\\
y_{n+1}\end{array}\right)=\left(\begin{array}{cc} 1-\varepsilon & \varepsilon\\
0 & 1\end{array}\right)\left(\begin{array}{c}f(x_{n+1})\\
f(y_{n+1})
\end{array}\right),
\end{equation}
where $x_{n}$ is the variable of the node being influenced and
$y_{n}$ the trajectory of the synchronized neighbourhood. Note
that it is implied in the equation that the input is a
trajectory of the uncoupled map ($y_{n+1}=f(y_{n})$). If this
were not the case we could not perform this analysis.

The eigenvalues of the linear operator are $\sigma_1=1$
($\textbf{e}_1=(1,1)$) and $\sigma_2=1-\varepsilon$
($\textbf{e}_2=(1,0)$). Therefore, the influenced node will
synchronize to its input if
\begin{equation}
\varepsilon>1-e^{-\lambda}.
\end{equation}

\section{Stability of the polysynchronous states}
\label{app:poly}

We study here the stability of different polysynchronous states in
the simplest structure capable of showing polysynchrony: the triplet
with transposition symmetry. In this case polysynchrony means full
synchronization of nodes $i$ and $k$. Thus, the quotient system of
the triplet is a completely connected pair and the possible
polysynchronous dynamics are therefore attractors of the completely
connected pair.

\subsection{Fixed point polysynchronous state}

\begin{figure}[h]
\begin{center}
\includegraphics[width=8cm]{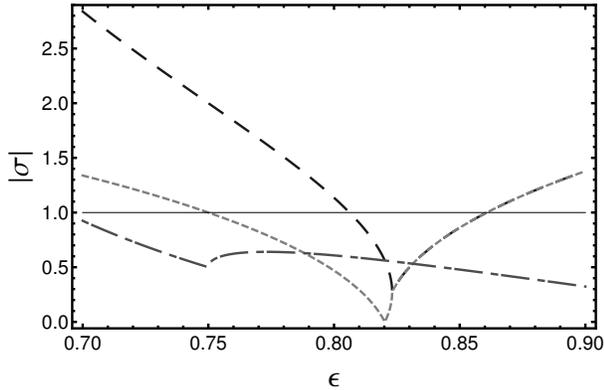}
\caption{\label{fig:eigenFPPoly}Eigenvalues of $J$ evaluated at the
fixed point $(c_1,c_2,c_1)$ as a function of $\varepsilon$. The three eigenvalues have modulus less than unity in the range $0.806186\lesssim\varepsilon\lesssim0.86$.}
\end{center}
\end{figure}

The completely connected pair has two fixed points $(c_1,c_2)$ and
$(c_2,c_1)$ with
\begin{eqnarray*}
c_1&=& \frac{1}{8(2\varepsilon-1)}(8\varepsilon-3+\sqrt{9-4\varepsilon(9-8\varepsilon)}),\\
c_2&=& \frac{1}{8(2\varepsilon-1)}(8\varepsilon-3-\sqrt{9-4\varepsilon(9-8\varepsilon)}),\\
\end{eqnarray*}
that are stable in the range
$0.806186\lesssim\varepsilon\lesssim0.86$. These allow the triplet
to have two possible polysynchronous fixed point states:
$(c_1,c_2,c_1)$ or $(c_2,c_1,c_2)$. The stability of this states can
be evaluated as the stability of a fixed point of a three
dimensional system by studying the absolute value of the eigenvalues
of the jacobian matrix at the fixed point. The jacobian matrix for
the triplet with transposition symmetry reads
\begin{widetext}
\begin{equation}
J(x^{i}, x^{j}, x^{k})=\left(
  \begin{array}{ccc}
    (1-\varepsilon)f'(x^i) & \varepsilon f'(x^j) & 0  \\
    \varepsilon\frac{m'}{m}f'(x^i) & (1-\varepsilon)f'(x^j) & \varepsilon\frac{m-m'}{m}f'(x^k)  \\
    0 & \varepsilon f'(x^j) & (1-\varepsilon)f'(x^k)
      \end{array}
\right), \nonumber
\end{equation}
\end{widetext}
In Fig.~\ref{fig:eigenFPPoly} we represent the absolute value of
the eigenvalues of $J(c_1,c_2,c_1)$ as a function of $\varepsilon$
and we can clearly see that the polysynchronous state is stable in
all the stability range of the fixed points.

It is very interesting to note that the eigenvalues are independent
of $m'$ and therefore our conclusions are also valid for a
completely connected pair with an outgoing link to a third node (as
in the 3-node subsystem of Fig.~\ref{fig:three}.\textbf{e.}).

\subsection{Period-2 polysynchronous state}

As described in \cite{gyllenberg1993does}, the period-2 orbit of the
completely connected pair has as its elements
\begin{eqnarray*}
(z_1,z_2)&=&(\frac{u+v+1}{2},\frac{u-v+1}{2}),\\
u&:=&\frac{1}{4g},\\
v&:=&\frac{\sqrt{8g^2-2g-1}}{4g},\\
g&:=&1-2\varepsilon.
\end{eqnarray*}
This period-2 dynamics is stable in the range
$0.13925\lesssim\varepsilon\lesssim0.193814$.

Similarly to the fixed-point case, the stability of the period-2
polysychronous state can be studied as the stability of a period-2
orbit of a three dimensional system. Therefore, we should observe
the eigenvalues of the jacobian matrix
\begin{equation}
J_2(z_1,z_2)=J(z_1,z_2,z_1)\cdot J(z_2,z_1,z_2).
\end{equation}

\begin{figure}[h]
\begin{center}
\includegraphics[width=8cm]{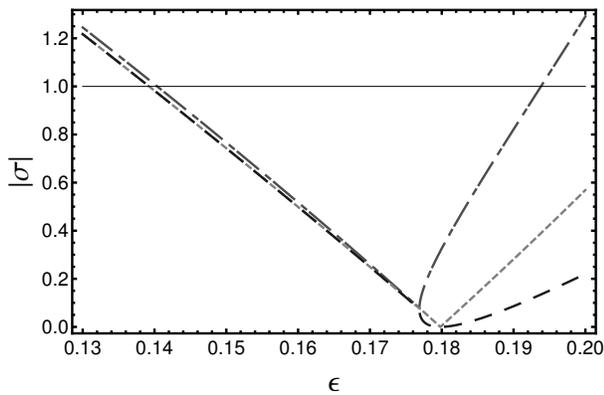}
\caption{Eigenvalues of $J_2$ as a function of
$\varepsilon$. The three eigenvalues have modulus less than unity in the range $0.140375\lesssim\varepsilon\lesssim0.193814$. \label{fig:eigenP2Poly}}
\end{center}
\end{figure}

The eigenvalues of $J_2$ have been graphed in Fig.~\ref{fig:eigenP2Poly}
as a function of the coupling strength. From this figure we see that
the period-2 polysynchronous state is stable when
$0.140375\lesssim\varepsilon\lesssim0.193814$.

Interestingly, and contrary to the fixed-point case, there is an
interval of coupling strengths
($0.13925\lesssim\varepsilon\lesssim0.140375$) where the period-2
orbit is stable in the pair but the period-2 polysynchronous state
of the triplet is unstable. It is in this interval of values where the spatial period-doubling phenomena appears (see Fig.~\ref{fig:openflow} and description in the text).

\subsection{Chaotic polysynchronous state}

To study the stability of the chaotic polysynchronous states we
start by making the following change of variable
\begin{eqnarray*}
U&=&\frac{x^i+x^k}{2},\\
V&=&x^j,\\
W&=&\frac{x^k-x^i}{2}.
\end{eqnarray*}

In these new variables, the dynamics is given by
\begin{widetext}
\begin{eqnarray*}
U_{n+1}&=&\frac{1}{2}[(1-\varepsilon)(f(U_n-W_n)+f(U_n+W_n))]+\varepsilon f(V_n),\\
V_{n+1}&=&\varepsilon\frac{m'}{m} f(U_n-W_n)+(1-\varepsilon)f(V_n)+\varepsilon\frac{m'-m}{m}f(U_n+V_n),\\
W_{n+1}&=&\frac{(1-\varepsilon)}{2}[f(U_n+W_n)-f(U_n-W_n)].
\end{eqnarray*}
\end{widetext}

The polysynchronous chaotic state corresponds to the case $W_n=0$
with $U_{n}$ and $V_{n}$ following a non-synchronous chaotic orbit
that we denote $U^\ast$,$V^\ast$. By expanding the equation for
$W_{n+1}$ around $(U^\ast,V^\ast,0)$ we obtain
\begin{equation}
W_{n+1}\approx-2(1-\varepsilon)U^\ast_nW_n.
\end{equation}

Assuming ergodicity, the transverse Lyapunov exponent (transverse to
the surface of $\mathbb{R}^3$ where the polysynchronous orbit lies)
can be written as \cite{pikovsky2003synchronization}
\begin{equation}\label{eq:Lyapunov}
\lambda_{\top}=\ln|-2(1-\varepsilon)|+\lambda^{\ast},
\end{equation}

where $\lambda^{\ast}$ is the average Lyapunov exponent of the orbit
$U^{\ast}_n$ and has the upper bound $\ln 2$, that would correspond
to the complete synchronization of $U_n$ and $V_n$ (or equivalently,
$x^i$ and $x^j$). We need $\lambda_{\top}<0$ for the polysynchronous
chaotic state to be stable. This condition provides us with a
relation between the Lyapunov exponent of the chaotic orbit of
$U^{\ast}$ and the minimum coupling strength necessary for
polysynchrony to be stable
\begin{equation}
\varepsilon_{min}=1-\frac{e^{-\lambda^{\ast}}}{2}.
\end{equation}

It is easy to see that $\varepsilon_{min}$ is always greater than
0.5 for $\lambda^{\ast}\le\ln 2$. This is in agreement with the
numerical experiments, which do not witness chaotic polysynchrony
for $\varepsilon<0.5$.
\begin{figure}[h]
\begin{center}
\includegraphics[width=6cm]{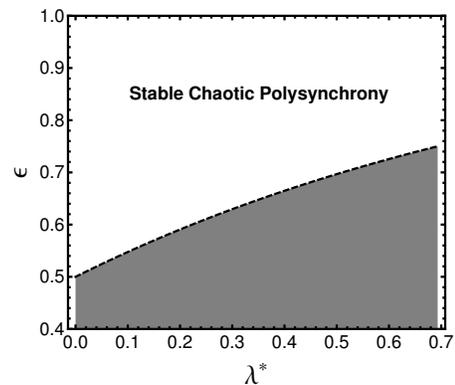}
\caption{\label{fig:chaoticPoly}Region of stability of the chaotic
polysynchronous state as a function of the Lyapunov exponent of the
chaotic orbit. The dashed line corresponds to $\varepsilon_{min}$.}
\end{center}
\end{figure}

\medskip
\begin{acknowledgments}
PG is partially funded by the UK EPSRC grants EP/E050441/1 (CICADA) and EP/I01912X/1 (EPSRC Mathematics Platform Engagement grant). VBS is
partially supported by contract MICINN (AYA2010-22111-C03-02). VBS also thanks Dept. F\'isica Te\`orica (Universitat de Val\`encia) for hospitality.
\end{acknowledgments}

\bibliography{Adapt_ms_long}

\begin{thebibliography}{42}%
\makeatletter
\providecommand \@ifxundefined [1]{%
 \@ifx{#1\undefined}
}%
\providecommand \@ifnum [1]{%
 \ifnum #1\expandafter \@firstoftwo
 \else \expandafter \@secondoftwo
 \fi
}%
\providecommand \@ifx [1]{%
 \ifx #1\expandafter \@firstoftwo
 \else \expandafter \@secondoftwo
 \fi
}%
\providecommand \natexlab [1]{#1}%
\providecommand \enquote  [1]{``#1''}%
\providecommand \bibnamefont  [1]{#1}%
\providecommand \bibfnamefont [1]{#1}%
\providecommand \citenamefont [1]{#1}%
\providecommand \href@noop [0]{\@secondoftwo}%
\providecommand \href [0]{\begingroup \@sanitize@url \@href}%
\providecommand \@href[1]{\@@startlink{#1}\@@href}%
\providecommand \@@href[1]{\endgroup#1\@@endlink}%
\providecommand \@sanitize@url [0]{\catcode `\\12\catcode `\$12\catcode
  `\&12\catcode `\#12\catcode `\^12\catcode `\_12\catcode `\%12\relax}%
\providecommand \@@startlink[1]{}%
\providecommand \@@endlink[0]{}%
\providecommand \url  [0]{\begingroup\@sanitize@url \@url }%
\providecommand \@url [1]{\endgroup\@href {#1}{\urlprefix }}%
\providecommand \urlprefix  [0]{URL }%
\providecommand \Eprint [0]{\href }%
\providecommand \doibase [0]{http://dx.doi.org/}%
\providecommand \selectlanguage [0]{\@gobble}%
\providecommand \bibinfo  [0]{\@secondoftwo}%
\providecommand \bibfield  [0]{\@secondoftwo}%
\providecommand \translation [1]{[#1]}%
\providecommand \BibitemOpen [0]{}%
\providecommand \bibitemStop [0]{}%
\providecommand \bibitemNoStop [0]{.\EOS\space}%
\providecommand \EOS [0]{\spacefactor3000\relax}%
\providecommand \BibitemShut  [1]{\csname bibitem#1\endcsname}%
\let\auto@bib@innerbib\@empty
\bibitem [{\citenamefont {Boccaletti}\ \emph {et~al.}(2006)\citenamefont
  {Boccaletti}, \citenamefont {Latora}, \citenamefont {Moreno}, \citenamefont
  {Chavez},\ and\ \citenamefont {Hwang}}]{boccaletti2006complex}%
  \BibitemOpen
  \bibfield  {author} {\bibinfo {author} {\bibfnamefont {S.}~\bibnamefont
  {Boccaletti}}, \bibinfo {author} {\bibfnamefont {V.}~\bibnamefont {Latora}},
  \bibinfo {author} {\bibfnamefont {Y.}~\bibnamefont {Moreno}}, \bibinfo
  {author} {\bibfnamefont {M.}~\bibnamefont {Chavez}}, \ and\ \bibinfo {author}
  {\bibfnamefont {D.}~\bibnamefont {Hwang}},\ }\href@noop {} {\bibfield
  {journal} {\bibinfo  {journal} {Phys. Rep.}\ }\textbf {\bibinfo {volume}
  {424}},\ \bibinfo {pages} {175} (\bibinfo {year} {2006})}\BibitemShut
  {NoStop}%
\bibitem [{\citenamefont {Newman}\ \emph {et~al.}(2006)\citenamefont {Newman},
  \citenamefont {Barabasi},\ and\ \citenamefont {Watts}}]{newman2006structure}%
  \BibitemOpen
  \bibfield  {author} {\bibinfo {author} {\bibfnamefont {M.}~\bibnamefont
  {Newman}}, \bibinfo {author} {\bibfnamefont {A.}~\bibnamefont {Barabasi}}, \
  and\ \bibinfo {author} {\bibfnamefont {D.}~\bibnamefont {Watts}},\
  }\href@noop {} {\emph {\bibinfo {title} {{The Structure and Dynamics of
  Networks}}}}\ (\bibinfo  {publisher} {Princeton University Press},\ \bibinfo
  {year} {2006})\BibitemShut {NoStop}%
\bibitem [{\citenamefont {Gross}\ and\ \citenamefont
  {Blasius}(2008)}]{Gross2008}%
  \BibitemOpen
  \bibfield  {author} {\bibinfo {author} {\bibfnamefont {T.}~\bibnamefont
  {Gross}}\ and\ \bibinfo {author} {\bibfnamefont {B.}~\bibnamefont
  {Blasius}},\ }\href@noop {} {\bibfield  {journal} {\bibinfo  {journal} {J. R.
  Soc. Interface}\ }\textbf {\bibinfo {volume} {5}},\ \bibinfo {pages} {259}
  (\bibinfo {year} {2008})}\BibitemShut {NoStop}%
\bibitem [{\citenamefont {Gross}\ and\ \citenamefont
  {Sayama}(2009)}]{gross2009adaptive}%
  \BibitemOpen
  \bibfield  {author} {\bibinfo {author} {\bibfnamefont {T.}~\bibnamefont
  {Gross}}\ and\ \bibinfo {author} {\bibfnamefont {H.}~\bibnamefont {Sayama}},\
  }\href@noop {} {\emph {\bibinfo {title} {{Adaptive Networks: Theory, Models
  and Applications}}}}\ (\bibinfo  {publisher} {Springer Verlag},\ \bibinfo
  {year} {2009})\BibitemShut {NoStop}%
\bibitem [{\citenamefont {Ito}\ and\ \citenamefont {Kaneko}(2001)}]{Ito2001}%
  \BibitemOpen
  \bibfield  {author} {\bibinfo {author} {\bibfnamefont {J.}~\bibnamefont
  {Ito}}\ and\ \bibinfo {author} {\bibfnamefont {K.}~\bibnamefont {Kaneko}},\
  }\href@noop {} {\bibfield  {journal} {\bibinfo  {journal} {Phys. Rev. Lett.}\
  }\textbf {\bibinfo {volume} {88}},\ \bibinfo {pages} {028701} (\bibinfo
  {year} {2001})}\BibitemShut {NoStop}%
\bibitem [{\citenamefont {Ito}\ and\ \citenamefont {Kaneko}(2003)}]{Ito2003}%
  \BibitemOpen
  \bibfield  {author} {\bibinfo {author} {\bibfnamefont {J.}~\bibnamefont
  {Ito}}\ and\ \bibinfo {author} {\bibfnamefont {K.}~\bibnamefont {Kaneko}},\
  }\href@noop {} {\bibfield  {journal} {\bibinfo  {journal} {Phys. Rev. E}\
  }\textbf {\bibinfo {volume} {67}},\ \bibinfo {pages} {046226} (\bibinfo
  {year} {2003})}\BibitemShut {NoStop}%
\bibitem [{\citenamefont {Fan}\ and\ \citenamefont
  {Chen}(2004)}]{fan2004evolving}%
  \BibitemOpen
  \bibfield  {author} {\bibinfo {author} {\bibfnamefont {Z.}~\bibnamefont
  {Fan}}\ and\ \bibinfo {author} {\bibfnamefont {G.}~\bibnamefont {Chen}},\
  }\href@noop {} {\bibfield  {journal} {\bibinfo  {journal} {Int. J. Mod. Phys.
  B}\ }\textbf {\bibinfo {volume} {18}},\ \bibinfo {pages} {2540} (\bibinfo
  {year} {2004})}\BibitemShut {NoStop}%
\bibitem [{\citenamefont {Berg}\ and\ \citenamefont
  {Leeuwen}(2004)}]{Berg2004}%
  \BibitemOpen
  \bibfield  {author} {\bibinfo {author} {\bibfnamefont {D.~V.~D.}\
  \bibnamefont {Berg}}\ and\ \bibinfo {author} {\bibfnamefont {C.~V.}\
  \bibnamefont {Leeuwen}},\ }\href@noop {} {\bibfield  {journal} {\bibinfo
  {journal} {Europhys. Lett.}\ }\textbf {\bibinfo {volume} {65}},\ \bibinfo
  {pages} {459} (\bibinfo {year} {2004})}\BibitemShut {NoStop}%
\bibitem [{\citenamefont {Gong}\ and\ \citenamefont
  {Leeuwen}(2004)}]{Gong2004}%
  \BibitemOpen
  \bibfield  {author} {\bibinfo {author} {\bibfnamefont {P.}~\bibnamefont
  {Gong}}\ and\ \bibinfo {author} {\bibfnamefont {C.~V.}\ \bibnamefont
  {Leeuwen}},\ }\href@noop {} {\bibfield  {journal} {\bibinfo  {journal}
  {Europhys. Lett.}\ }\textbf {\bibinfo {volume} {67}},\ \bibinfo {pages} {328}
  (\bibinfo {year} {2004})}\BibitemShut {NoStop}%
\bibitem [{\citenamefont {Zhou}\ and\ \citenamefont {Kurths}(2006)}]{Zhou2006}%
  \BibitemOpen
  \bibfield  {author} {\bibinfo {author} {\bibfnamefont {C.}~\bibnamefont
  {Zhou}}\ and\ \bibinfo {author} {\bibfnamefont {J.}~\bibnamefont {Kurths}},\
  }\href@noop {} {\bibfield  {journal} {\bibinfo  {journal} {Phys. Rev. Lett.}\
  }\textbf {\bibinfo {volume} {96}},\ \bibinfo {pages} {164102} (\bibinfo
  {year} {2006})}\BibitemShut {NoStop}%
\bibitem [{\citenamefont {Lu}(2007)}]{lu2007adaptive}%
  \BibitemOpen
  \bibfield  {author} {\bibinfo {author} {\bibfnamefont {W.}~\bibnamefont
  {Lu}},\ }\href@noop {} {\bibfield  {journal} {\bibinfo  {journal} {Chaos}\
  }\textbf {\bibinfo {volume} {17}},\ \bibinfo {pages} {023122} (\bibinfo
  {year} {2007})}\BibitemShut {NoStop}%
\bibitem [{\citenamefont {Aoki}\ and\ \citenamefont
  {Aoyagi}(2009)}]{aoki2009co}%
  \BibitemOpen
  \bibfield  {author} {\bibinfo {author} {\bibfnamefont {T.}~\bibnamefont
  {Aoki}}\ and\ \bibinfo {author} {\bibfnamefont {T.}~\bibnamefont {Aoyagi}},\
  }\href@noop {} {\bibfield  {journal} {\bibinfo  {journal} {Phys. Rev. Lett.}\
  }\textbf {\bibinfo {volume} {102}},\ \bibinfo {pages} {034101} (\bibinfo
  {year} {2009})}\BibitemShut {NoStop}%
\bibitem [{\citenamefont {Li}\ \emph {et~al.}(2010)\citenamefont {Li},
  \citenamefont {Guan},\ and\ \citenamefont {Lai}}]{Li2010}%
  \BibitemOpen
  \bibfield  {author} {\bibinfo {author} {\bibfnamefont {M.}~\bibnamefont
  {Li}}, \bibinfo {author} {\bibfnamefont {S.}~\bibnamefont {Guan}}, \ and\
  \bibinfo {author} {\bibfnamefont {C.-H.}\ \bibnamefont {Lai}},\ }\href@noop
  {} {\bibfield  {journal} {\bibinfo  {journal} {New J. Phys.}\ }\textbf
  {\bibinfo {volume} {12}},\ \bibinfo {pages} {103032} (\bibinfo {year}
  {2010})}\BibitemShut {NoStop}%
\bibitem [{\citenamefont {Kwok}\ \emph {et~al.}(2007)\citenamefont {Kwok},
  \citenamefont {Jurica}, \citenamefont {Raffone},\ and\ \citenamefont
  {Van~Leeuwen}}]{kwok2007robust}%
  \BibitemOpen
  \bibfield  {author} {\bibinfo {author} {\bibfnamefont {H.}~\bibnamefont
  {Kwok}}, \bibinfo {author} {\bibfnamefont {P.}~\bibnamefont {Jurica}},
  \bibinfo {author} {\bibfnamefont {A.}~\bibnamefont {Raffone}}, \ and\
  \bibinfo {author} {\bibfnamefont {C.}~\bibnamefont {Van~Leeuwen}},\
  }\href@noop {} {\bibfield  {journal} {\bibinfo  {journal} {Cogn. Neurodyn.}\
  }\textbf {\bibinfo {volume} {1}},\ \bibinfo {pages} {39} (\bibinfo {year}
  {2007})}\BibitemShut {NoStop}%
\bibitem [{\citenamefont {Meisel}\ and\ \citenamefont
  {Gross}(2009)}]{meisel2009adaptive}%
  \BibitemOpen
  \bibfield  {author} {\bibinfo {author} {\bibfnamefont {C.}~\bibnamefont
  {Meisel}}\ and\ \bibinfo {author} {\bibfnamefont {T.}~\bibnamefont {Gross}},\
  }\href@noop {} {\bibfield  {journal} {\bibinfo  {journal} {Phys. Rev. E}\
  }\textbf {\bibinfo {volume} {80}},\ \bibinfo {pages} {061917} (\bibinfo
  {year} {2009})}\BibitemShut {NoStop}%
\bibitem [{\citenamefont {Gomez~Portillo}\ \emph {et~al.}(2009)\citenamefont
  {Gomez~Portillo}, \citenamefont {Gleiser},\ and\ \citenamefont
  {Sporns}}]{gomez2009adaptive}%
  \BibitemOpen
  \bibfield  {author} {\bibinfo {author} {\bibfnamefont {I.}~\bibnamefont
  {Gomez~Portillo}}, \bibinfo {author} {\bibfnamefont {P.}~\bibnamefont
  {Gleiser}}, \ and\ \bibinfo {author} {\bibfnamefont {O.}~\bibnamefont
  {Sporns}},\ }\href@noop {} {\bibfield  {journal} {\bibinfo  {journal} {PLoS
  One}\ }\textbf {\bibinfo {volume} {4}},\ \bibinfo {pages} {418} (\bibinfo
  {year} {2009})}\BibitemShut {NoStop}%
\bibitem [{\citenamefont {Gleiser}\ and\ \citenamefont
  {Spoormaker}(2010)}]{gleiser2010modelling}%
  \BibitemOpen
  \bibfield  {author} {\bibinfo {author} {\bibfnamefont {P.}~\bibnamefont
  {Gleiser}}\ and\ \bibinfo {author} {\bibfnamefont {V.}~\bibnamefont
  {Spoormaker}},\ }\href@noop {} {\bibfield  {journal} {\bibinfo  {journal}
  {Philos. T. R. Soc. A}\ }\textbf {\bibinfo {volume} {368}},\ \bibinfo {pages}
  {5633} (\bibinfo {year} {2010})}\BibitemShut {NoStop}%
\bibitem [{\citenamefont {Gross}\ \emph {et~al.}(2006)\citenamefont {Gross},
  \citenamefont {D'Lima},\ and\ \citenamefont {Blasius}}]{Gross2006}%
  \BibitemOpen
  \bibfield  {author} {\bibinfo {author} {\bibfnamefont {T.}~\bibnamefont
  {Gross}}, \bibinfo {author} {\bibfnamefont {C.}~\bibnamefont {D'Lima}}, \
  and\ \bibinfo {author} {\bibfnamefont {B.}~\bibnamefont {Blasius}},\
  }\href@noop {} {\bibfield  {journal} {\bibinfo  {journal} {Phys. Rev. Lett.}\
  }\textbf {\bibinfo {volume} {96}},\ \bibinfo {pages} {208701} (\bibinfo
  {year} {2006})}\BibitemShut {NoStop}%
\bibitem [{\citenamefont {Gross}\ and\ \citenamefont
  {Kevrekidis}(2008)}]{gross2008robust}%
  \BibitemOpen
  \bibfield  {author} {\bibinfo {author} {\bibfnamefont {T.}~\bibnamefont
  {Gross}}\ and\ \bibinfo {author} {\bibfnamefont {I.}~\bibnamefont
  {Kevrekidis}},\ }\href@noop {} {\bibfield  {journal} {\bibinfo  {journal}
  {Europhys. Lett.}\ }\textbf {\bibinfo {volume} {82}},\ \bibinfo {pages}
  {38004} (\bibinfo {year} {2008})}\BibitemShut {NoStop}%
\bibitem [{\citenamefont {Shaw}\ and\ \citenamefont
  {Schwartz}(2010)}]{shaw2010enhanced}%
  \BibitemOpen
  \bibfield  {author} {\bibinfo {author} {\bibfnamefont {L.}~\bibnamefont
  {Shaw}}\ and\ \bibinfo {author} {\bibfnamefont {I.}~\bibnamefont
  {Schwartz}},\ }\href@noop {} {\bibfield  {journal} {\bibinfo  {journal}
  {Phys. Rev. E}\ }\textbf {\bibinfo {volume} {81}},\ \bibinfo {pages} {046120}
  (\bibinfo {year} {2010})}\BibitemShut {NoStop}%
\bibitem [{\citenamefont {Kozma}\ and\ \citenamefont
  {Barrat}(2008)}]{kozma2008consensus}%
  \BibitemOpen
  \bibfield  {author} {\bibinfo {author} {\bibfnamefont {B.}~\bibnamefont
  {Kozma}}\ and\ \bibinfo {author} {\bibfnamefont {A.}~\bibnamefont {Barrat}},\
  }\href@noop {} {\bibfield  {journal} {\bibinfo  {journal} {Phys. Rev. E}\
  }\textbf {\bibinfo {volume} {77}},\ \bibinfo {pages} {016102} (\bibinfo
  {year} {2008})}\BibitemShut {NoStop}%
\bibitem [{\citenamefont {Nardini}\ \emph {et~al.}(2008)\citenamefont
  {Nardini}, \citenamefont {Kozma},\ and\ \citenamefont
  {Barrat}}]{nardini2008s}%
  \BibitemOpen
  \bibfield  {author} {\bibinfo {author} {\bibfnamefont {C.}~\bibnamefont
  {Nardini}}, \bibinfo {author} {\bibfnamefont {B.}~\bibnamefont {Kozma}}, \
  and\ \bibinfo {author} {\bibfnamefont {A.}~\bibnamefont {Barrat}},\
  }\href@noop {} {\bibfield  {journal} {\bibinfo  {journal} {Phys. Rev. Lett.}\
  }\textbf {\bibinfo {volume} {100}},\ \bibinfo {pages} {158701} (\bibinfo
  {year} {2008})}\BibitemShut {NoStop}%
\bibitem [{\citenamefont {Botella-Soler}\ and\ \citenamefont
  {Glendinning}(2012)}]{bg2011polyletter}%
  \BibitemOpen
  \bibfield  {author} {\bibinfo {author} {\bibfnamefont {V.}~\bibnamefont
  {Botella-Soler}}\ and\ \bibinfo {author} {\bibfnamefont {P.}~\bibnamefont
  {Glendinning}},\ }\href@noop {} {\bibfield  {journal} {\bibinfo  {journal}
  {Europhys. Lett.}\ }\textbf {\bibinfo {volume} {97}},\ \bibinfo {pages}
  {50004} (\bibinfo {year} {2012})}\BibitemShut {NoStop}%
\bibitem [{\citenamefont {Stewart}\ \emph {et~al.}(2003)\citenamefont
  {Stewart}, \citenamefont {Golubitsky},\ and\ \citenamefont
  {Pivato}}]{stewart2003symmetry}%
  \BibitemOpen
  \bibfield  {author} {\bibinfo {author} {\bibfnamefont {I.}~\bibnamefont
  {Stewart}}, \bibinfo {author} {\bibfnamefont {M.}~\bibnamefont {Golubitsky}},
  \ and\ \bibinfo {author} {\bibfnamefont {M.}~\bibnamefont {Pivato}},\
  }\href@noop {} {\bibfield  {journal} {\bibinfo  {journal} {SIAM J. Appl.
  Dynam. Sys.}\ }\textbf {\bibinfo {volume} {2}},\ \bibinfo {pages} {609}
  (\bibinfo {year} {2003})}\BibitemShut {NoStop}%
\bibitem [{\citenamefont {Golubitsky}\ \emph {et~al.}(2004)\citenamefont
  {Golubitsky}, \citenamefont {Nicol},\ and\ \citenamefont
  {Stewart}}]{golubitsky2004some}%
  \BibitemOpen
  \bibfield  {author} {\bibinfo {author} {\bibfnamefont {M.}~\bibnamefont
  {Golubitsky}}, \bibinfo {author} {\bibfnamefont {M.}~\bibnamefont {Nicol}}, \
  and\ \bibinfo {author} {\bibfnamefont {I.}~\bibnamefont {Stewart}},\
  }\href@noop {} {\bibfield  {journal} {\bibinfo  {journal} {J. Nonlinear
  Sci.}\ }\textbf {\bibinfo {volume} {14}},\ \bibinfo {pages} {207} (\bibinfo
  {year} {2004})}\BibitemShut {NoStop}%
\bibitem [{\citenamefont {Field}(2004)}]{field2004combinatorial}%
  \BibitemOpen
  \bibfield  {author} {\bibinfo {author} {\bibfnamefont {M.}~\bibnamefont
  {Field}},\ }\href@noop {} {\bibfield  {journal} {\bibinfo  {journal} {Dynam.
  Sys.}\ }\textbf {\bibinfo {volume} {19}},\ \bibinfo {pages} {217} (\bibinfo
  {year} {2004})}\BibitemShut {NoStop}%
\bibitem [{\citenamefont {Aguiar}\ \emph {et~al.}(2009)\citenamefont {Aguiar},
  \citenamefont {Ashwin}, \citenamefont {Dias},\ and\ \citenamefont
  {Field}}]{aguiar2009dynamics}%
  \BibitemOpen
  \bibfield  {author} {\bibinfo {author} {\bibfnamefont {M.}~\bibnamefont
  {Aguiar}}, \bibinfo {author} {\bibfnamefont {P.}~\bibnamefont {Ashwin}},
  \bibinfo {author} {\bibfnamefont {A.}~\bibnamefont {Dias}}, \ and\ \bibinfo
  {author} {\bibfnamefont {M.}~\bibnamefont {Field}},\ }\href@noop {}
  {\bibfield  {journal} {\bibinfo  {journal} {J. Nonlinear Sci.}\ ,\ \bibinfo
  {pages} {1}} (\bibinfo {year} {2009})}\BibitemShut {NoStop}%
\bibitem [{\citenamefont {Agarwal}\ and\ \citenamefont
  {Field}(2010)}]{agarwal2010dynamical}%
  \BibitemOpen
  \bibfield  {author} {\bibinfo {author} {\bibfnamefont {N.}~\bibnamefont
  {Agarwal}}\ and\ \bibinfo {author} {\bibfnamefont {M.}~\bibnamefont
  {Field}},\ }\href@noop {} {\bibfield  {journal} {\bibinfo  {journal}
  {Nonlinearity}\ }\textbf {\bibinfo {volume} {23}},\ \bibinfo {pages} {1245}
  (\bibinfo {year} {2010})}\BibitemShut {NoStop}%
\bibitem [{\citenamefont {Kestler}\ \emph {et~al.}(2007)\citenamefont
  {Kestler}, \citenamefont {Kinzel},\ and\ \citenamefont
  {Kanter}}]{Kestler2007}%
  \BibitemOpen
  \bibfield  {author} {\bibinfo {author} {\bibfnamefont {J.}~\bibnamefont
  {Kestler}}, \bibinfo {author} {\bibfnamefont {W.}~\bibnamefont {Kinzel}}, \
  and\ \bibinfo {author} {\bibfnamefont {I.}~\bibnamefont {Kanter}},\
  }\href@noop {} {\bibfield  {journal} {\bibinfo  {journal} {Phys. Rev. E}\
  }\textbf {\bibinfo {volume} {76}},\ \bibinfo {pages} {035202} (\bibinfo
  {year} {2007})}\BibitemShut {NoStop}%
\bibitem [{\citenamefont {Kestler}\ \emph {et~al.}(2008)\citenamefont
  {Kestler}, \citenamefont {Kopelowitz}, \citenamefont {Kanter},\ and\
  \citenamefont {Kinzel}}]{Kestler2008}%
  \BibitemOpen
  \bibfield  {author} {\bibinfo {author} {\bibfnamefont {J.}~\bibnamefont
  {Kestler}}, \bibinfo {author} {\bibfnamefont {E.}~\bibnamefont {Kopelowitz}},
  \bibinfo {author} {\bibfnamefont {I.}~\bibnamefont {Kanter}}, \ and\ \bibinfo
  {author} {\bibfnamefont {W.}~\bibnamefont {Kinzel}},\ }\href@noop {}
  {\bibfield  {journal} {\bibinfo  {journal} {Phys. Rev. E}\ }\textbf {\bibinfo
  {volume} {77}},\ \bibinfo {pages} {046209} (\bibinfo {year}
  {2008})}\BibitemShut {NoStop}%
\bibitem [{\citenamefont {Kanter}\ \emph {et~al.}(2011)\citenamefont {Kanter},
  \citenamefont {Zigzag}, \citenamefont {Englert}, \citenamefont {Geissler},\
  and\ \citenamefont {Kinzel}}]{Kanter2011}%
  \BibitemOpen
  \bibfield  {author} {\bibinfo {author} {\bibfnamefont {I.}~\bibnamefont
  {Kanter}}, \bibinfo {author} {\bibfnamefont {M.}~\bibnamefont {Zigzag}},
  \bibinfo {author} {\bibfnamefont {A.}~\bibnamefont {Englert}}, \bibinfo
  {author} {\bibfnamefont {F.}~\bibnamefont {Geissler}}, \ and\ \bibinfo
  {author} {\bibfnamefont {W.}~\bibnamefont {Kinzel}},\ }\href@noop {}
  {\bibfield  {journal} {\bibinfo  {journal} {Europhys. Lett.}\ }\textbf
  {\bibinfo {volume} {93}},\ \bibinfo {pages} {60003} (\bibinfo {year}
  {2011})}\BibitemShut {NoStop}%
\bibitem [{\citenamefont {Kaneko}(1990)}]{kaneko1990supertransients}%
  \BibitemOpen
  \bibfield  {author} {\bibinfo {author} {\bibfnamefont {K.}~\bibnamefont
  {Kaneko}},\ }\href@noop {} {\bibfield  {journal} {\bibinfo  {journal} {Phys.
  Lett. A}\ }\textbf {\bibinfo {volume} {149}},\ \bibinfo {pages} {105}
  (\bibinfo {year} {1990})}\BibitemShut {NoStop}%
\bibitem [{\citenamefont {Quince}\ \emph {et~al.}(2005)\citenamefont {Quince},
  \citenamefont {Higgs},\ and\ \citenamefont {McKane}}]{quince2005topological}%
  \BibitemOpen
  \bibfield  {author} {\bibinfo {author} {\bibfnamefont {C.}~\bibnamefont
  {Quince}}, \bibinfo {author} {\bibfnamefont {P.}~\bibnamefont {Higgs}}, \
  and\ \bibinfo {author} {\bibfnamefont {A.}~\bibnamefont {McKane}},\
  }\href@noop {} {\bibfield  {journal} {\bibinfo  {journal} {Ecol. Model.}\
  }\textbf {\bibinfo {volume} {187}},\ \bibinfo {pages} {389} (\bibinfo {year}
  {2005})}\BibitemShut {NoStop}%
\bibitem [{\citenamefont {Kaneko}(1985)}]{kaneko1985spatial}%
  \BibitemOpen
  \bibfield  {author} {\bibinfo {author} {\bibfnamefont {K.}~\bibnamefont
  {Kaneko}},\ }\href@noop {} {\bibfield  {journal} {\bibinfo  {journal} {Phys.
  Lett. A}\ }\textbf {\bibinfo {volume} {111}},\ \bibinfo {pages} {321}
  (\bibinfo {year} {1985})}\BibitemShut {NoStop}%
\bibitem [{\citenamefont {Willeboordse}\ and\ \citenamefont
  {Kaneko}(1995)}]{willeboordse1995pattern}%
  \BibitemOpen
  \bibfield  {author} {\bibinfo {author} {\bibfnamefont {F.}~\bibnamefont
  {Willeboordse}}\ and\ \bibinfo {author} {\bibfnamefont {K.}~\bibnamefont
  {Kaneko}},\ }\href@noop {} {\bibfield  {journal} {\bibinfo  {journal}
  {Physica D}\ }\textbf {\bibinfo {volume} {86}},\ \bibinfo {pages} {428}
  (\bibinfo {year} {1995})}\BibitemShut {NoStop}%
\bibitem [{\citenamefont {Rudzick}\ and\ \citenamefont
  {Pikovsky}(1996)}]{rudzick1996unidirectionally}%
  \BibitemOpen
  \bibfield  {author} {\bibinfo {author} {\bibfnamefont {O.}~\bibnamefont
  {Rudzick}}\ and\ \bibinfo {author} {\bibfnamefont {A.}~\bibnamefont
  {Pikovsky}},\ }\href@noop {} {\bibfield  {journal} {\bibinfo  {journal}
  {Phys. Rev. E}\ }\textbf {\bibinfo {volume} {54}},\ \bibinfo {pages} {5107}
  (\bibinfo {year} {1996})}\BibitemShut {NoStop}%
\bibitem [{\citenamefont {Yamaguchi}(1997)}]{yamaguchi1997mechanism}%
  \BibitemOpen
  \bibfield  {author} {\bibinfo {author} {\bibfnamefont {A.}~\bibnamefont
  {Yamaguchi}},\ }\href@noop {} {\bibfield  {journal} {\bibinfo  {journal}
  {Int. J. Bifurcat. Chaos}\ }\textbf {\bibinfo {volume} {7}},\ \bibinfo
  {pages} {1529} (\bibinfo {year} {1997})}\BibitemShut {NoStop}%
\bibitem [{\citenamefont {Parthasarathy}\ and\ \citenamefont
  {Guemez}(1998)}]{parthasarathy1998synchronisation}%
  \BibitemOpen
  \bibfield  {author} {\bibinfo {author} {\bibfnamefont {S.}~\bibnamefont
  {Parthasarathy}}\ and\ \bibinfo {author} {\bibfnamefont {J.}~\bibnamefont
  {Guemez}},\ }\href@noop {} {\bibfield  {journal} {\bibinfo  {journal} {Ecol.
  model.}\ }\textbf {\bibinfo {volume} {106}},\ \bibinfo {pages} {17} (\bibinfo
  {year} {1998})}\BibitemShut {NoStop}%
\bibitem [{\citenamefont {Pikovsky}\ \emph {et~al.}(2003)\citenamefont
  {Pikovsky}, \citenamefont {Rosenblum},\ and\ \citenamefont
  {Kurths}}]{pikovsky2003synchronization}%
  \BibitemOpen
  \bibfield  {author} {\bibinfo {author} {\bibfnamefont {A.}~\bibnamefont
  {Pikovsky}}, \bibinfo {author} {\bibfnamefont {M.}~\bibnamefont {Rosenblum}},
  \ and\ \bibinfo {author} {\bibfnamefont {J.}~\bibnamefont {Kurths}},\
  }\href@noop {} {\emph {\bibinfo {title} {{Synchronization: A universal
  concept in nonlinear sciences}}}}\ (\bibinfo  {publisher} {Cambridge Univ.
  Pr.},\ \bibinfo {year} {2003})\BibitemShut {NoStop}%
\bibitem [{\citenamefont {Gyllenberg}\ \emph {et~al.}(1993)\citenamefont
  {Gyllenberg}, \citenamefont {S\"oderbacka},\ and\ \citenamefont
  {Ericsson}}]{gyllenberg1993does}%
  \BibitemOpen
  \bibfield  {author} {\bibinfo {author} {\bibfnamefont {M.}~\bibnamefont
  {Gyllenberg}}, \bibinfo {author} {\bibfnamefont {G.}~\bibnamefont
  {S\"oderbacka}}, \ and\ \bibinfo {author} {\bibfnamefont {S.}~\bibnamefont
  {Ericsson}},\ }\href@noop {} {\bibfield  {journal} {\bibinfo  {journal}
  {Math. Biosci.}\ }\textbf {\bibinfo {volume} {118}},\ \bibinfo {pages} {25}
  (\bibinfo {year} {1993})}\BibitemShut {NoStop}%
\bibitem [{\citenamefont {Lloyd}(1995)}]{lloyd1995coupled}%
  \BibitemOpen
  \bibfield  {author} {\bibinfo {author} {\bibfnamefont {A.}~\bibnamefont
  {Lloyd}},\ }\href@noop {} {\bibfield  {journal} {\bibinfo  {journal} {J.
  Theor. Biol.}\ }\textbf {\bibinfo {volume} {173}},\ \bibinfo {pages} {217}
  (\bibinfo {year} {1995})}\BibitemShut {NoStop}%
\bibitem [{\citenamefont {Kendall}\ and\ \citenamefont
  {Fox}(1998)}]{kendall1998spatial}%
  \BibitemOpen
  \bibfield  {author} {\bibinfo {author} {\bibfnamefont {B.}~\bibnamefont
  {Kendall}}\ and\ \bibinfo {author} {\bibfnamefont {G.}~\bibnamefont {Fox}},\
  }\href@noop {} {\bibfield  {journal} {\bibinfo  {journal} {Theor. Popul.
  Biol.}\ }\textbf {\bibinfo {volume} {54}},\ \bibinfo {pages} {11} (\bibinfo
  {year} {1998})}\BibitemShut {NoStop}%
\end{thebibliography}%

\end{document}